\documentclass[a4paper,11pt]{article}
\usepackage{tablefootnote}
\usepackage[english]{babel}
\usepackage[T1]{fontenc}
\usepackage[flushmargin]{footmisc}
\usepackage{setspace}
\usepackage{float}
\usepackage{amsmath}
\usepackage{amsfonts}
\usepackage{amssymb}
\usepackage{ae}
\usepackage{placeins}
\usepackage[comma]{natbib}
\usepackage{caption}
\usepackage{adjustbox}
\usepackage{graphicx}
\usepackage{subcaption}
\usepackage{changepage}
\usepackage{tabularx,booktabs}

\newcolumntype{Y}{>{\centering\arraybackslash}X}
\usepackage[utf8]{inputenc}

\usepackage[a4paper,left=2.5cm,right=2.5cm,top=3cm,bottom=3cm]{geometry}
\usepackage[colorinlistoftodos]{todonotes}
\usepackage[colorlinks=true, allcolors=blue]{hyperref}
\begin{document} \doublespacing \pagestyle{plain}

	\begin{center}

		{\LARGE How causal machine learning can leverage marketing strategies: Assessing and improving the performance of a coupon campaign}

{\large
	\vspace{1.3cm}}

{\large Henrika Langen and Martin Huber}\medskip

{\normalsize University of Fribourg, Dept.\ of Economics \bigskip }
\end{center}

\bigskip

\noindent \textbf{Abstract:} {\small
	We apply causal machine learning algorithms to assess the causal effect of a marketing intervention, namely a coupon campaign, on the sales of a retailer. Besides assessing the average impacts of different types of coupons, we also investigate the heterogeneity of causal effects across different subgroups of customers, e.g.,\ between clients with relatively high vs.\ low prior purchases. Finally, we use optimal policy learning to determine (in a data-driven way) which customer groups should be targeted by the coupon campaign in order to maximize the marketing intervention's effectiveness in terms of sales. We find that only two out of the five coupon categories examined, namely coupons applicable to the product categories of drugstore items and other food, have a statistically significant positive effect on retailer sales. The assessment of group average treatment effects reveals substantial differences in the impact of coupon provision across customer groups, particularly across customer groups as defined by prior purchases at the store, with drugstore coupons being particularly effective among customers with high prior purchases and other food coupons among customers with low prior purchases. Our study provides a use case for the application of causal machine learning in business analytics to evaluate the causal impact of specific firm policies (like marketing campaigns) for decision support.
}

{\small \smallskip }



\bigskip
{\small \smallskip {\scriptsize \noindent
}  }

{\thispagestyle{empty}\pagebreak  }

{\small \renewcommand{\thefootnote}{\arabic{footnote}} %
	\setcounter{footnote}{0}  \pagebreak \setcounter{footnote}{0} \pagebreak %
	\setcounter{page}{1} }

\section{Introduction}\label{intro}

Over the last two decades, the amount of customer data available to marketers has increased dramatically with new data types such as social media, clickstream, search query and supermarket scanner data on the rise. The increasing availability of customer Big Data has spawned a new stream of literature on machine learning (ML) methods and tools in the field of business and marketing. The ML literature on designing marketing campaigns ranges from research on modeling customer behavior (e.g. \cite*{Xiaetal2019}, \cite*{Huetal2019}), price sensitivity (e.g. \cite*{Arevalillo2021}) and purchase decisions (e.g. \cite*{Donnellyetal2021}) to studies on the development of personalized product recommendation systems (e.g. \cite*{Ramzanetal2019}, \cite*{AnithaKalaiarasu2021}), customer churn management (e.g. \cite*{GordiniVeglio2017}) and acquisition of new customers (e.g.\cite*{Luketal2019}).

A common feature of these studies is that they are based on predictive ML, i.e., on identifying patterns of variables in the data in order to use them for predicting an outcome of interest (e.g., sales). This is done by training predictive models in one part of the data and determining the best performing model (with the smallest possible prediction error) in the other part of the data. Under some commonly used ML algorithms,  the identified model serves as a black box, i.e., it is based on functions that are too complex for any human to understand (as in so-called deep learning), while in other cases, the model has an explicit  (and thus comprehensible) structure. In any case, however, such predictive ML models generally do  not provide insights into the causal effects of specific variables or interventions (such as a marketing campaign) on the outcome of interest. Thus, predictive ML, although appropriate for making educated guesses about outcomes based on certain patterns observed in the data, is not well suited for determining and comparing the effectiveness of possible courses of action, which would be relevant for decision support, e.g.\ for optimally designing a marketing campaign.\footnote{To predict an outcome of interest based on predictor variables, ML aims at minimizing the prediction error by optimally trading off prediction bias and variance. When multiple variables capture the same relevant predictive feature, i.e., are correlated with that feature, ML algorithms may identify some of these variables as relevant predictors while attaching little importance to others, regardless of the variables' causal effect on the outcome. For instance, variables that do not directly or only modestly affect the outcome may enter the prediction model as relevant predictors, simply  because they are correlated with other variables that actually affect the outcome. For this reason, it may happen that these other variables play little or no role in the predictive model, even though they have a causal impact on the outcome, simply because they provide little additional information for the prediction. Therefore, predictive ML is generally not suitable for the causal analysis of ‘what if’ questions, such as how a change in a coupon campaign strategy will affect customer behavior.}

To improve on the shortcomings of predictive ML in evaluating the impact of implementing vs.\ not implementing a specific intervention, a fast growing literature in econometrics and statistics has been developing so-called causal ML algorithms. In this paper, we demonstrate the application of such methods in the context of business analytics for decision support, that is, for evaluating a marketing intervention. More precisely, we make use of the so-called causal forest approach by \cite*{AtheyTibshiraniWager2019} to assess the causal effect of marketing campaigns, in which customers were provided coupons for different product types, on customers' purchasing behavior, i.e. the difference in their expected behavior with and without being targeted by a coupon campaign. While predictive ML algorithms are not able to isolate the causal effects of coupons on customers' purchasing behavior from the influence of background characteristics (e.g.\ socio-economic characteristics and price sensitivity) which jointly influence coupon reception and purchasing behavior, the causal forest approach can do so under certain assumptions.

One crucial condition is  that all variables that jointly affect coupon reception and purchasing behavior are observed in the data and can thus be controlled for. This condition is known as selection-on-observables or unconfoundedness assumption. Under further conditions on the quality of the ML models estimated as part of the causal forest approach for predicting purchasing behavior and coupon reception as a function of the observed variables, the causal forest approach permits evaluating the mean impact of the coupons on all customers, as well as across specific subgroups or customer segments (e.g.\ different age groups). Our results suggest, for instance, a positive overall effect of coupons for drugstore items. For coupons applicable to ready-to-eat food as well as meat and seafood, on the other hand, we do not find a statistically significant overall effect. An analysis of the effect of drugstore coupons across different customer subgroups reveals that these coupons particularly affect customers with high pre-campaign spending as well as low- to middle-income customers.

Furthermore, we apply optimal policy learning based on ML as proposed by \cite*{AtheyWager2021}, in order to learn from the data which customer segments should be optimally targeted by coupon campaigns such that the overall (or net after cost) effect is maximized. In contrast to predictive ML, optimal policy learning allows, under certain conditions, identifying the coupon provision policy which is most effective in terms of its impact on sales. This is done by first assessing the expected effects in different customer segments and then selecting those segments as target groups in which the effects are sufficiently high. The estimated optimal policy for coupons applicable to meat and seafood, for instance, suggests that such coupons should be issued to low-income customers whose pre-campaign spending did not exceed a certain level, to middle-to-high-income customers aged 46 years or older who purchased something from the store in the period prior to the campaign, as well as to  middle-to-high-income households with at least five members who did not purchase anything from the store in the pre-campaign period.

The paper proceeds as follows: Section \ref{motivation} outlines the current state of quantitative research in the marketing literature and motivates the application of causal ML methods in the field of marketing.  
Section \ref{data} introduces and describes the retailer's sales data to be analyzed. Section \ref{identification} defines the causal effects of interest based on so-called counterfactual reasoning, discusses the conditions  required for applying causal ML (such as the selection-on-observables assumption) and describes the algorithms for causal analysis and optimal policy learning. Section \ref{results} provides the results of the evaluation of the retailer's coupon campaigns as well as the optimal coupon allocation. Section \ref{conclusion} concludes.

\section{Motivation}\label{motivation}

The evaluation of the causal impact of discount campaigns plays a significant role in the earlier marketing literature from the `pre-Big-Data era', see e.g. \cite*{InmanMcAlister1994}, \cite*{Rajuetal1994}, \cite*{LeoneSrinivasan1996} and \cite*{Krishnaetl1999} for studies on causal effects of coupon provision. However, the last two decades have seen a surge of predictive ML applications in business analytics, which appear to increasingly dominate causal analysis in marketing as well. In a keyword-search-based literature review, \cite*{Marianietal2021} find that the number of publications on predictive ML and Artificial Intelligence (AI) in marketing, consumer research and psychology has grown exponentially in the past decade (2010-21). The systematic literature reviews by \cite*{Mustaketal2021} and \cite*{MaSun2020} paint a similar picture, with the latter stating that the rise of ML in marketing began with applications of support vector machines, a specific type of ML algorithm. This was then followed by studies that introduced text analysis, topic modelling and reinforcement learning into marketing research, as well as by marketing applications of deep learning, and network embedding. Questions about the impact of marketing campaigns, the influence of certain external factors on the success of a campaign and the heterogeneity of campaign effects across customer segments appeared to become of comparatively less importance (see e.g. \cite*{HairSarstedt2021}, \cite*{MaSun2020}), even though most recently, the marketing literature saw first applications of causal ML alogithms (such as causal trees).

The following sections summarize the current state of research on discount campaigns using causal inference (Section \ref{causallit}) and predictive ML (Section \ref{predMLlit}). This serves as the basis for motivating the use of causal ML to evaluate and optimize discount campaigns and to approach various other marketing and business decisions in Section \ref{causalMLmot}.

\subsection{Causal Inference in Marketing}\label{causallit}

A number of studies assess the causal effects of specific marketing campaigns on consumer response to the campaigns. These studies typically rely on (field) experiments or traditional methods for causal inference based on observational data. In the latter case, researchers must assume that all variables that jointly affect the intervention and purchasing behavior are observed in the data and can thus be controlled for. \cite*{RubinRichard2006} apply propensity score matching to evaluate the effect of marketing interventions aimed at physicians in order to promote the prescription of a `lifestyle' drug. They also rank the physicians according to their estimated expected individual-level effects, which in turn can be used to derive a tailored marketing strategy. \cite*{ReimersXie2019} assess the effect of e-coupon provision on alcohol sales by means of a difference-in-differences approach, exploiting the fact that the restaurants in their sample issued e-coupons at different points in time. See also \cite*{Xingetal2020}, \cite*{Halvorsenetal2016} and \cite*{Zhangetal2017} for further examples of observational and experimental studies examining the effect of coupon provision or other discount campaigns on consumer behavior.

Other contributions analyze the heterogeneity of marketing effects across customer characteristics and the circumstances under which customers are targeted by coupon and other promotional campaigns. Among them, \cite*{GopalakrishnanPark2021} investigate whether high- and low-consumption customers, as defined by their purchasing behavior during the 12 months prior to the experiment, differ in their responsiveness to coupon campaigns. \cite*{Andrewsetal2016} study whether the level of occupancy (or crowdedness) of a subway affects passengers' response to mobile advertising campaigns  and find a statistically significant positive association. Based on a field experiment, \cite*{Spiekermannetal2011} conclude that proximity to the location for which coupons are distributed influences coupon redemption, and that this association is much more pronounced in the city center than in suburban areas.

Furthermore, several studies evaluate how certain configurations of coupons, such as face value, distribution method and expiry date, affect consumer behavior. The experimental studies by \cite*{Zhengetal2021} and \cite*{Biswasetal2013} assess how the size of discounts affects consumers' perceptions of product quality and purchase intentions. \cite*{LeoneSrinivasan1996} use supermarket scanner data to analyze the effect of coupon face value on sales and profits, while \cite*{AndersonSimester2004} study the long-term effects of discount size on the purchasing behavior of new and established customers in an experimental setting. Other contributions as e.g.\  \cite*{GopalakrishnanPark2021}, \cite*{Jiaetal2018}, \cite*{ChoiCoulter2012}, \cite*{Krishnaetl1999} and \cite*{InmanMcAlister1994} analyze how further aspects of coupon and discount campaign design affect consumer behavior.

\subsection{Predictive ML in Marketing}\label{predMLlit}

In recent years, many studies have focused on ML-based prediction of coupon redemption and associated sales. They use ML algorithms to model customer behavior as a function of customers' previous transactions, their response to past coupon/discount campaigns and their socio-economic characteristics in order to predict the likelihood of customers to redeem coupons or take up discounts and make purchases.

\cite*{PusztovaBabic2020} and \cite*{HeJiang2017} compare the performance of different ML-based classification algorithms in predicting coupon redemption in digital marketing campaigns. The first study concludes that so-called Support Vector Machines provide the most accurate predictions, while the latter study shows that the gradient boosting framework `XGBoost' performs best. \cite*{Greensteinetal2017}  introduce an  algorithm  that combines co-clustering and random forest classification to predict redemption of mobile restaurant coupons based on demographic and contextual variables such as the consumer's distance to the restaurant relative to the size of the coupon discount. \cite*{Renetal2021} developped a two-stage model for estimating the probability of coupon redemption, consisting of a first stage in which customers are clustered based on their past purchase and redemption behavior, followed by a second stage of fitting prediction models for the different customer clusters. Furthermore, several studies such as \cite*{Koehnetal2020}, \cite*{Xiaoetal2021} and \cite*{Zhengetal2021} predict customer behavior in the context of coupon or other discount campaigns by means of several ML methods.

\subsection{Causal ML in Marketing}\label{causalMLmot}

Under certain conditions like the selection-on-observables assumption, implying that all variables that jointly affect the intervention and purchasing behavior are observed in the data and can thus be controlled for, causal ML methods allow for the evaluation of causal effects of coupon/discount campaigns as well as effect heterogeneity across customer segments. In contrast to more traditional methods of causal inference, they can leverage the full amount of information available to marketers, which may be large in the era of `Big Data'. That is, causal ML can address research questions such as those described in Section \ref{causallit} based on high-dimensional observational data containing a large set of background variables that could serve as control variables. Examples include socio-economic characteristics of customers, geographic or time-related information, weather, economic circumstances, and many more. Causal ML is based on combining causal inference approaches with ML algorithms for data-driven selection of control variables when estimating causal effects and/or their heterogeneity across customer segments.

The rise of predictive ML has prompted e.g.\ \cite*{Anderson2008}, \cite*{Lycett2013} and \cite*{Erevellesetal2016} to argue that theory-based causal inference has lost some of its relevance for business decisions in light of the large datasets and sophisticated predictive ML methods available to marketers today. However, these views were soon challenged in several studies that emphasize the importance of causal reasoning and risks of basing decisions based solely on correlations, see e.g.\ \cite*{CowlsSchroeder2015} and \cite*{GolderMacy2014}. In more recent years, a growing number of contributions have further stressed the importance of integrating ML and causal inference, see e.g. \cite*{HairSarstedt2021}. Among them, \cite*{Hunermundetal2021}, who investigate the use of causal methods in business analytics by combining qualitative interviews and quantitative surveys among data scientists and managers in a mixed-methods research design. They document an ongoing shift in corporate decision making away from an exclusive focus on predictive ML and towards the use of causal methods, based on both observational and experimental data.

Yet, to date, applications of causal ML to marketing research appear to be relatively scarce, with the exception of large tech companies operating in the field of social media or online commerce. To the best of our knowledge, there are virtually no studies that evaluate the causal effect of coupon campaigns on customer behavior using causal ML, as we do in this paper. \cite*{Smithetal2021} use predictive ML for deriving optimal coupon targeting strategies and estimate the profits that would accrue under those strategies out of sample, i.e.\ in parts of the data not used for deriving the strategies. The profits are estimated based on the potential outcomes framework, which is also the basis of causal ML. However, the study by \cite*{Smithetal2021} is conceptually different from ours in that it uses the potential outcomes framework to compare coupon targeting strategies inferred from different predictive ML algorithms, while we apply an algorithm based on the potential outcomes framework (namely the optimal policy learning approach of \cite*{AtheyWager2021}) to derive a coupon targeting strategy.

One study in the field of marketing which does consider causal ML is \cite*{Gordonetal2022}. They assess the performance of so-called Double Machine Learning (DML), see \cite*{Chetal2018}, and propensity score matching, see \cite*{RosenbaumRubin1983}, for estimating the causal effect of conversion ads on Facebook. Such ads aim to increase online activity like page visits, sales and views on an external website. For their analysis, the authors take advantage of the fact that Facebook offers businesses the opportunity to assess their ad campaigns by means of randomized experiments. \cite*{Gordonetal2022} compare the effect estimates based on DML and propensity score matching with those from the experiments, finding that DML outperforms propensity score matching, but that both approaches overestimate the effect substantially. This highlights the importance of observing and appropriately controlling for all factors jointly affecting the intervention and customer behavior when causally assessing marketing interventions. Also, \cite*{Huberetal2021} consider DML when analyzing observational data to investigate whether discounted tickets induce Swiss railway customers to reschedule their journeys, e.g.\ to shift demand away from peak hours. 

\cite*{Narangetal2019} apply causal forests, the causal ML framework developed by \cite*{WagerAthey2018} and \cite*{AtheyTibshiraniWager2019} also used in this study (see Section \ref{methodscausalML}), to assess the heterogeneity across shoppers in how mobile app failures affect the frequency, volume, and monetary value of their purchases.  \cite*{Guoetal2021} assess the effect of a law requiring pharmaceutical firms to disclose their marketing payments to physicians on the firms' payments to physicians using a Difference-in-Differences approach and assess expected individual-level effect heterogeneity by means of causal forests.  \cite*{ZhangLuo2021} incorporate causal forests in their study on modelling restaurant survival as a function of photos posted on social networks. They find that the total volume of user-generated content and the extent to which user photos are rated as helpful have a significant positive effect on the likelihood of restaurant survival. Another study from the broader field of marketing that uses causal ML is \cite*{Cagalaetal2021}. The authors apply causal ML to determine the strategy for distributing gifts among potential donors to a fundraising campaign that maximizes expected net donations. They find that the identified optimal targeting rule outperforms different non-targeted gift distribution rules, even when the optimal targeting rule is estimated based only on publicly available geographic information or on data from a previous fundraising campaign conducted in a similar sample.

In the following, we will use coupon promotions as a running example to highlight the merits of causal ML in business analytics and marketing research. In the context of coupon campaigning strategies, marketers are arguably interested not only in predicting customer behavior, but also in measuring the causal effects of alternative campaigns on customer behavior. Such effects correspond to the difference in the customers' (average) behavior when being vs.\ not being addressed by a particular campaign. Intuitively, this requires comparing a customer's observed behavior under the actual assigned coupon with the potential (and not directly observed) behavior that would have occurred had coupon provision been different from that actually observed, an approach commonly referred to as counterfactual reasoning. Such a causal assessment is necessary for determining whether and to which extent a campaign is effective in altering customer behavior and for understanding how customer behavior would change if coupons were distributed differently.

In a predictive ML model, however, the predictive power of coupon provision on customer behavior generally does not correspond to such a causal effect, because it is affected by so-called regularization bias, i.e., a bias that arises in the context of ML algorithms shrinking the importance of certain predictors in order to reduce the variance of the prediction and thereby improve the overall predictive performance. Regularization bias may occur, for instance, when coupon provision is strongly correlated with other (good) predictors (such as previous purchases) and/or when its effect on consumer behavior is comparably small, so that coupon provision has little predictive value. A further issue is selection bias, meaning that coupons may pick up the effects of other variables  whose importance has been shrunk by the ML algorithm. The implementation of coupon campaigns should be based on estimations of causal effects that avoid regularization and selection bias, as is the case with causal ML algorithms such as DML and causal forests.

The necessity of estimating the causal effect of coupon campaigns, rather than merely predicting customer behavior, can be illustrated by means of a simple example. Suppose a retailer estimates a model to predict sales based on observational data from a previous coupon campaign in which (in an attempt to re-activate dormant customers) coupons were  distributed primarily among customers who had not been in the store for a while, rather than among frequent shoppers. The estimated predictive model might indicate a negative association between coupon provision and sales, since the coupon campaign is likely to re-activate only some inactive customers, so that  the (formerly) inactive customers on average purchase less than the frequent shoppers. The true effect of receiving a coupon, however, might actually be positive. A positive effect implies that when comparing two groups of (formerly) inactive customers with comparable background characteristics (like willingness to buy), where the first receives coupons while the second does not, the average purchases of the first group are higher. The predictive model therefore confuses (or confounds) the causal effect of the coupon campaign with that of being a  dormant vs.\ a frequent shopper, thus incorrectly pointing to a negative effect.

In a second scenario, the retailer decides to issue coupons in the store. This way, frequent shoppers are regularly provided with coupons, while dormant customers rarely if ever receive any. A predictive model now detects a positive relationship between the provision of coupons and sales, although the actual effect of providing coupons could be negative, namely if frequent customers use the coupons for products they would have bought anyway. If the campaigns were evaluated using predictive methods and the results were misinterpreted as causal, marketers would come to the conclusion that the first campaign was ineffective while  the second was effective. Causal methods, on the other hand, enable marketers under certain conditions to control for such biases, in our example due to differences in purchasing behavior between frequent and dormant customers, and to consistently estimate the effect of coupon campaigns. Further, these methods can also be applied to assess effect heterogeneity and identify an optimal coupon distribution scheme (or policy) that targets those customers whose average purchases are sufficiently responsive to receiving a coupon.

In causal studies on discounts, the impact of providing coupons is typically assessed either based on random experiments or observational data from previous campaigns, controlling for observed characteristics or covariates that are likely to be associated with both coupon provision and consumer behavior. Conventional, i.e.,\ non-ML-based, causal inference methods require the researcher or analyst to manually select covariates based on theoretical considerations, domain knowledge, intuition and/or previous empirical findings. Examples for such covariates in the context of campaign evaluations include past purchasing behavior, exposure to previous campaigns, and socio-economic characteristics such as age, gender, or income. Manual selection of covariates entails the risk of omitting important control variables and may even be practically infeasible in Big Data contexts with a very large set of potential covariates (e.g.,\ collected from online platforms), including unstructured data containing, e.g.,\ text or clickstreams. Furthermore, conventional causal inference methods require the researcher to specify how, i.e., through which functional form (like, e.g.,\ a linear model), the selected covariates are associated with coupon provision and purchasing behavior. Causal ML methods, in contrast, permit taking advantage of the full amount of information in the data to detect relevant covariates (which have an important influence on coupon provision and consumer behavior) in a data-driven way and control for them, as well as to flexibly estimate the functional form of statistical associations. Still, the observational data have to meet certain conditions, as described in Section \ref{identifyingassumptions}.

The argument for counterfactual reasoning made further above also applies to efforts of optimizing the distribution of coupons across segments of customers, i.e.,\ optimal policy learning, as discussed, e.g.,\ in \cite*{Manski2004}, \cite*{HiranoPorter2008}, \cite*{Stoye2009}, and \cite*{KitagawaTetenov2018}. Basing optimization on predictive ML models, as advocated in several studies on predicting coupon redemption (e.g. \cite*{Koehnetal2020}, \cite*{Renetal2021}, \cite*{Greensteinetal2017}), ignores the fact that predictive models do generally not provide information on causal effects and their heterogeneity across different customer segments. Causal ML-based policy learning as suggested by \cite*{AtheyWager2021}, on the other hand, is a causal ML approach to inferring allocation schemes which ensure that those customers for whom sufficiently large effects can be expected are targeted by the campaign. In our empirical application, we demonstrate how causal ML methods can help evaluate coupon campaigns and support marketing-related decision making. We analyze customer data from a retail store and first evaluate the average effect of providing customers with a coupon (of a certain type) on the monetary value of their purchases.  In a second step, we demonstrate how optimal policy learning can be used for detecting customer segments that should or should not be targeted by coupon campaigns to maximize the effectiveness of these campaigns.

\section{Data}\label{data}

In our empirical application, we analyze sales data on coupon campaigns of a retailer, which are available on the data science platform \cite*{kaggle2021} under the denomination `Predicting Coupon Redemption'. The dataset contains information on socio-economic characteristics of retail store customers, the coupons they have received during the campaigns as well as on their coupon redemption and purchasing behavior. The retail store ran several campaigns issuing coupons with discounts for certain products, with some coupons being applicable to individual products only and others to a range of products. In each of the 18 partially time-overlapping campaigns falling into the time span covered by the dataset, the store distributed 1 to 208 different coupon types each applicable to up to 12,000 products, most of which belong to the same product category. The coupons were distributed in such a way that each customer received 0 to 37 different coupons per campaign with the composition of this set of coupons varying between the recipients. Apart from the information on provided coupons, the dataset contains details on all purchases made by each registered customer between January 2012 and July 2013, including the date of the transaction, the redeemed coupons, the product type of each purchased product and the price paid.

For our analysis, we group the coupons into five broad categories mirroring the products they can be used for. More concisely, we distinguish between coupons applicable for ready-to-eat food items, meat and seafood, other food, drugstore items and other non-food products\footnote{The coupons of each category are applicable to the following product categories defined by the retailer: (a) ready-to-eat food coupons: `Bakery', `Restaurant', `Prepared Food', `Dairy, Juices \& Snacks', (b) meat and seafood coupons: `Meat', `Packaged Meat', `Seafood', (c) coupons applicable to other food: `Grocery', `Salads', `Vegetables (cut)', `Natural Products', (d) drugstore coupons: `Pharmaceutical', `Skin \& Hair Care', and (e) coupons applicable to other non-food products: `Flowers \& Plants', `Garden', `Travel', `Miscellaneous'}. One could arguably also be interested in more fine-grained coupon categories or in a paricular coupon or discount type rather than in our broader coupon categories, which would, however, require a larger dataset to obtain satisfactory statistical power. Due to the temporal overlap of campaign periods, we need to redefine them such that each of the resulting artificially generated campaign periods coincides with the validity period of a given set of coupons. That is, all coupons which are valid in some artificial campaign period are valid during the entire period. By doing so, we can fully attribute changes in purchasing behavior from one artificial campaign period to another to the coupons valid in the respective periods. From now on, the 33 newly defined artificial campaign periods will simply be referred to as campaign periods. To account for differences in the duration of campaign periods, we consider the average per-day expenditures per customer and campaign period as our outcome of interest. For estimating the causal effect of coupon provision on the buying behavior, we pool the customer-specific purchases across campaign periods, yielding 33 observations per customer.

\begin{table}
	\captionsetup{font=footnotesize}
	\begin{singlespace}
		\centering
		\adjustbox{width = 0.9\textwidth, center}{%
			
			\begin{tabular}{lccccc}
				\toprule
				variable & Overall & Coupon Receivers & Non-Receivers & Diff & p-val \\
				N & 50,624 & 15,327 & 35,297 & &\\
				\midrule
				daily expenditures & 202 & 245 & 184 & 61 & 0 \\
				age: & & & & & \\
				\ \ \ \ \ 18-25 & 0.028 & 0.031 & 0.027 & 0 & 0.02 \\
				\ \ \ \ \ 26-35 & 0.082 & 0.102 & 0.074 & 0.03 & 0 \\
				\ \ \ \ \ 36-45 & 0.118 & 0.141 & 0.108 & 0.03 & 0 \\
				\ \ \ \ \ 46-55 & 0.171 & 0.191 & 0.163 & 0.03 & 0 \\
				\ \ \ \ \ 56-70 & 0.037 & 0.045 & 0.034 & 0.01 & 0 \\
				\ \ \ \ \ 70+ & 0.043 & 0.039 & 0.045 & -0.01 & 0 \\
				\ \ \ \ \ unknown & 0.52 & 0.451 & 0.55 & -0.1 & 0 \\
				family size:  & & & & & \\
				\ \ \ \ \ 1 & 0.157 & 0.171 & 0.15 & 0.02 & 0 \\
				\ \ \ \ \ 2 & 0.192 & 0.213 & 0.182 & 0.03 & 0 \\
				\ \ \ \ \ 3 & 0.066 & 0.079 & 0.06 & 0.02 & 0 \\
				\ \ \ \ \ 4 & 0.03 & 0.04 & 0.026 & 0.01 & 0 \\
				\ \ \ \ \ 5+ & 0.036 & 0.047 & 0.031 & 0.02 & 0 \\
				\ \ \ \ \ unknown & 0.52 & 0.451 & 0.55 & -0.1 & 0 \\
				marital status: & & & & & \\
				\ \ \ \ \ married & 0.2 & 0.234 & 0.186 & 0.05 & 0 \\
				\ \ \ \ \ unmarried & 0.072 & 0.084 & 0.067  & 0.02 & 0 \\
				\ \ \ \ \ unknown & 0.728 & 0.682 & 0.747 & -0.07 & 0 \\
				dwelling type: & & & & & \\
				\ \ \ \ \ rented & 0.026 & 0.033 & 0.023 & 0.01 & 0 \\
				\ \ \ \ \ owned & 0.454 & 0.516 & 0.428 & 0.09 & 0 \\
				\ \ \ \ \ unknown & 0.52 & 0.451 & 0.55 & -0.1 & 0 \\
				income group: & & & & & \\
				\ \ \ \ \ 1 & 0.037 & 0.042 & 0.035 & 0.01 & 0 \\
				\ \ \ \ \ 2 & 0.043 & 0.051 & 0.04 & 0.01 & 0 \\
				\ \ \ \ \ 3 & 0.044 & 0.049 & 0.042 & 0.01 & 0 \\
				\ \ \ \ \ 4 & 0.104 & 0.113 & 0.1 & 0.01 & 0 \\
				\ \ \ \ \ 5 & 0.118 & 0.137 & 0.11 & 0.03 & 0 \\
				\ \ \ \ \ 6 & 0.056 & 0.061 & 0.053 & 0.01 & 0 \\
				\ \ \ \ \ 7 & 0.02 & 0.023 & 0.019 & 0 & 0.01 \\
				\ \ \ \ \ 8 & 0.023 & 0.03 & 0.021 & 0.01 & 0 \\
				\ \ \ \ \ 9 & 0.018 & 0.024 & 0.016 & 0.01 & 0 \\
				\ \ \ \ \ 10 & 0.006 & 0.006 & 0.006 & 0 & 0.89 \\
				\ \ \ \ \ 11 & 0.003 & 0.003 & 0.003 & 0 & 0.43 \\
				\ \ \ \ \ 12 & 0.006 & 0.01 & 0.005 & 0 & 0 \\
				\ \ \ \ \ unknown & 0.52 & 0.451 & 0.55 & -0.1 & 0 \\
				coupons redeemed &  & 0.030 & & &  \\
				\bottomrule
		\end{tabular}}
		\caption[Descriptive statistics]{Mean of the variables in the total sample ('Overall`), among coupon receivers and non-receivers as well as the mean difference across treatment states and the p-value of a two-sample t-test.}\label{table:desccoupon}
	\end{singlespace}
\end{table}

Table \ref{table:desccoupon} provides some descriptive statistics for our data, namely on observed customer characteristics, the share of coupons redeemed and daily in-store spanding (descriptive statistics on the composition of daily expenditures by product type are provided in Table \ref{tab:compositionexpenditures} in the appendix). The table reports the mean of these variables in the total sample of 50,624 observations, as well as among observations that received a coupon and among those that did not. Further, it contains the mean difference in these variables between coupon receivers and non-receivers, as well as the p-value of a two-sample t-test. In some 30\% of the observations, customers received at least one coupon. Furthermore, customers who received a coupon had on average higher expenditures in the retail store than customers who did not, suggesting that the retailer did not target its previous campaigns to re-activate dormant customers. We also see that the retailer does not have information on the socio-economic characteristics of all customers in the registry, but only for about half of them, as the corresponding variable values are unknown for many observations (see the coding `unknown'). Such a high rate of non-response in measuring variables can entail selection bias when estimating the effects of interest. For this reason, we will conduct several robustness checks in the empirical analysis to follow further below (see Section \ref{resultsrobustness}). The descriptive statistics also reveal that some socio-economic characteristics as well as their observability are correlated with the reception of coupons. For example, customers aged 70 years or older are less likely to be targeted by a coupon campaign. The main difference in the likelihood of receiving a coupon seems to be between customers whose socioeconomic characteristics are not available and those whose characteristics are known, with the former less likely to receive a coupon.

As is noted in several studies (e.g. \cite*{Danaheretal2015}, \cite*{Spiekermannetal2011}) coupon redemption rates are typically low, not exceeding 1 to 3\% on average. This is also the case in our data, as only in 3\% of the observations of coupon recipients did they actually redeem a coupon. However, as mentioned further above, coupons may not only influence customer behavior when redeemed, but may also serve as an advertising tool which attracts customers to the store even without them redeeming the coupon.

\section{Identification}\label{identification}

\subsection{Causal Effect}

We are interested in estimating the causal effect of a specific intervention, commonly referred to as `treatment' in causal analysis and henceforth denoted by $D$, on an outcome of interest, denoted by $Y$.\footnote{Throughout this paper, capital letters denote random variables and small letters specific values of random variables.} In our context, $D$ reflects the reception or non-reception of coupons and $Y$ the purchasing behavior, measured as the average per-day expenditures during the coupon validity period. In the simplest treatment definition, $D$ is binary and takes the value 1 when the respective customer is provided with a coupon and 0 if this is not the case. Mathematically speaking, the value $d$ which treatment $D$ can take satisfies $d \in \{0,1\}$. The set of observations with $d = 1$ is commonly referred to as the treatment group, those for which $d= 0$ are called control group. Our subsequent discussion of causal effects and the statistical assumptions required for their measurement will focus on this binary treatment case for the sake of simplicity. However, our empirical analysis will also separately consider the effects of receiving coupons for five product categories, by running separate estimations for the comparison of each category to not receiving any coupons. This implies that the assumptions introduced in Section \ref{identifyingassumptions} need to hold for each of these categories. For discussions of multi-valued treatments, see e.g.\ \cite*{Im00} and \cite*{Le01}.

For defining the causal effect of coupon provision, we rely on the potential outcome framework pioneered by \cite*{Neyman23} and \cite*{Rubin1974}. Let $Y(d)$ denote the potential (rather than observed) outcome under a specific treatment value $d \in \{0,1\}$. That is, $Y(1)$ corresponds to a customer's potential purchasing behavior if she received a coupon, while $Y(0)$ is the behavior without a coupon. The causal effect of the coupon thus corresponds to the difference in the purchasing behavior with and without coupon, $Y(1)-Y(0)$, but can unfortunately not be directly assessed for any customer. This is due to the impossibility of observing customers at the same point in time under two mutually exclusive coupon assignments (1 vs.\ 0), which is known as the `fundamental problem of causal inference', see \cite*{Holland86}.
This follows from the fact that the outcome $Y$ which is observed in the data corresponds to the potential purchasing behavior under the coupon assignment actually received, namely $Y=Y(1)$ for those receiving a coupon ($d=1$), and $Y(0)=Y$ for those who do not ($d=0$). For coupon recipients, however, $Y(0)$ cannot be observed in the data, while for customers without a coupon $Y(1)$ remains unknown.

Even though causal effects are fundamentally unidentifiable at the individual level, we may, under the assumptions outlined further below, evaluate them at more aggregate levels, i.e.,\ based on groups of treated and nontreated individuals. One causal parameter which is typically of crucial interest is the average causal effect, also known as average treatment effect (ATE), i.e.,\ the average effect of coupon assignment $D$ on purchasing behavior $Y$ among the total of customers. Formally, the ATE, which we henceforth denote by $\Delta$, corresponds to the difference in the average potential outcomes $Y(1)$ and $Y(0)$:
\begin{eqnarray}\label{ATE}
	\Delta=E[Y(1)-Y(0)],
\end{eqnarray}
where $`E[...]'$ stands for `expectation', which is simply the average in the population.

\subsection{Identifying Assumptions}\label{identifyingassumptions}
In order to identify the ATE defined in the previous section, we need to impose several identifying assumptions, which are outlined in this section. We note that in the subsequent discussion, `$\bot$' stands for statistical independence. Further, $X$ denotes the set of covariates that should not be affected by treatment $D$ and therefore be observed before or at, but not after, treatment. \\~
\textbf{Assumption 1 (conditional independence of the treatment):}\\~
$Y(d) \bot D | X$ for all $d  \in \{0,1\}$.
Assumption 1 states that the treatment is conditionally independent of the outcome when controlling for the covariates, and is known as `selection on observables', `unconfoundedness' or `ignorable treatment assignment', see e.g.\ \cite*{RosenbaumRubin1983}. The assumption implies that there are no unobservables jointly affecting the treatment assignment and the outcomes conditional on the covariates. This condition is satisfied if the coupons are quasi-randomly distributed among observations with the same values in $X$. The retailer may therefore base the distributing of coupons on customer or market characteristics observed in the data, however, not on unobserved characteristics that affect purchasing behavior even after controlling for the observed ones. 

We control for the variables in Table \ref{table:desccoupon}, period fixed effects, the customers' average daily pre-campaign spending by product category, as well as for the coupons she received and redeemed in the period prior to the campaign. When evaluating the effect of specific coupon categories, we also include dummies that indicate whether a customer received coupons from another category at the moment of treatment assignment. This is because the availability of other coupons influences purchase behavior and is likely to be correlated with the probability of receiving coupons of the category under study. The reason for including period fixed effects is that there is no information available on holidays or weekdays on which the store is closed or has shortened opening hours, that is, circumstances that may affect purchasing behavior. Also, the retailer is likely to distribute coupons differently across campaign periods.  Including pre-campaign expenditures allows controlling for general differences in purchasing behavior between customers that might be correlated with the likelihood of receiving coupons, since the retailer presumably bases decisions about whom to allocate which coupon(s) on past purchasing behavior.

The covariates considered in our estimation are similar to those included in studies on the effect of coupon campaigns that rely on traditional causal inference approaches, see, e.g., \cite{Xingetal2020} and \cite{Hsieh2010}, both of which control for some demographic characteristics as well as for a proxy for the customers' economic situation and their purchasing behavior before the coupon campaign under study. Unlike the methods used in these studies, however, the causal ML approach we apply in this study allows covariates to enter into the estimation in a flexible, possibly non-linear way, and does not require pre-selection of variables based on theoretical considerations.

Studies on predicting coupon redemption by means of ML mostly rely exclusively on observable customer behavior and coupon characteristics as predictors of coupon redemption while not including socio-demographic characteristics of customers, see, e.g., \cite*{Greensteinetal2017} use and \cite*{HeJiang2017}.

In their study on the performance of causal ML in evaluating Facebook ads,  \cite*{Gordonetal2022} include users' gender, age and household size but - unlike our data - their data lack information on users' economic situation, such as their income, employment status, or wealth. They also use several Facebook-specific covariates measuring users' activity on Facebook (likes, posts, type of device used and interests explicitely expressed on Facebook). Furthermore, they take into account users' response to earlier ads from other companies, which is comparable to the covariates on pre-campaign purchasing behavior, coupon reception and coupon redemption considered in our analysis. Despite the large differences in the amount of information available in the Facebook study and our analysis, we cannot conclude that the set of covariates in our estimation is insufficient.  For one, the algorithms used by Facebook to determine the target audience for ad placement are far more complex and information-hungry than a retailer's coupon strategy; and Facebook users' decision about whether or not to respond to a Facebook ad is likely to be complex and dependent on several of the characteristics considered in the algorithm (which is why they are considered in Facebook's ad placement algorithm). In order to successfully apply causal ML methods, the authors of the Facebook study had to take into account all the information that is incorporated in Facebook's ad placement algorithms, just as we need to consider the information based on which the retailer distributed its coupons, namely the information available in the customer database. \\~
\textbf{Assumption 2 (common support):}\\~
$0 < Pr(D = 1|X) < 1$.

Assumption 2 states that the conditional probability of being treated given $X$, in the following referred to as the treatment propensity score, is larger than zero and smaller than one. This so-called common support condition implies that for all values the covariates might take, customers have a non-zero chance of being treated and a non-zero chance of not being treated. While this assumption is imposed w.r.t.\ to the total of a (large) population, meaning that  both treated and non-treated customers exist conditional on $X$,  we can and should also verify it in the data. In our sample, common support appears to be satisfied, as there exist no combinations of covariate values for which either only customers with coupons (of a certain category) or no coupons exist. Appendix \ref{appendix:propensity} shows the distribution of the estimated propensity scores for receiving coupons (of a particular type) among recipients and non-recipients of that particular coupon(s). The distributions overlap (although the overlap is partially thin), i.e., for each observation in one group, observations can be found in the other group that are comparable with respect to the propensity score.

Another condition that needs to be satisfied is the so-called Stable Unit Treatment Value Assumption (SUTVA), see, e.g.,\ \cite*{Rubin1980}. In our context, SUTVA rules out that the coupons provided to one individual affect the potential outcome of another individual. The assumption that there are no inter-personal spillover effects of coupon campaigns  may be problematic in our setting.  Customers receiving coupons may  induce their peers to make purchases by, for instance, telling peers about the products they bought when redeeming the coupon or by visiting the store together with peers. On the other hand, customers with coupons may also redeem their coupons to buy the coupon-discounted products not only for themselves but also for their peers, thereby reducing the purchases made by their peers. Such scenarios appear particularly likely when there are several members of the same household in the customer base. There is ongoing research on how to deal with such SUTVA violations under certain assumptions like the observability of groups affected by spillovers, see e.g. \cite*{Sobel2006}, \cite*{HongRaudenbush2006}, \cite*{HudgensHalloran2008}  \cite*{TchetgenVanderWeele2012}, \cite*{AronowSamii2017}, \cite*{HuberSteinmayr2021} and  \cite*{Quetal2021}. However, in our dataset, the relationships between customers are not observable, meaning the data does not allow accounting for possible spillovers of providing coupons to one customer on the outcomes of other customers. If such spillovers existed in our case, they could entail an under- or overestimation of the effect of coupons on purchasing behavior, depending on whether the spillovers occur primarily through treated customers inducing non-treated peers to make purchases or through treated customers redeeming coupons to purchase products for their peers, with the former entailing an overestimation of the outcome under non-treatment and the latter leading to an underestimation.

SUTVA also requires that for every individual in the population, there is a single potential outcome value associated with each treatment state, meaning that there are no different versions of the coupons leading to different potential outcomes. In many empirical applications, it appears likely that at least some aspects of SUTVA are violated, and for this reason, there exist several relaxations of this assumption. In our case, the requirement that there be no different treatment versions is particularly problematic given that we group different coupons into broader categories. The treatment of being provided with coupon(s) from one category comprises the receipt of different coupons, each applicable to a distinct set of products from the respective product category. If a customer is not equally interested in all products belonging to that product category, the customer may only redeem a coupon and/or change her purchasing behavior if the coupon is applicable to certain products. For this reason, we are in a setting where there are different treatment versions, each possibly associated with a different potential outcome. 

\cite*{VanderWeeleetal2013} relax the original SUTVA by allowing for the existence of different unobservable versions of the treatment as long as there are no different versions of  non-treatment and the treatment versions are assigned randomly conditional on the covariates $X$. This permits assessing the average effects of certain bundles of coupons (rather than specific coupons as under the original SUTVA) vs.\ not receiving any coupons. Indeed, the assumption that there is only one version of non-treatment is satisfied in our analysis of the effect of receiving some vs.\ no  coupons, under the assumption that the marketer has not run any undocumented discount campaigns during the study period. Furthermore, when assessing the effects of coupons applicable to specific product categories, we control for all other coupons that each customer received at treatment assignment, which in turn creates non-treatment states that are necessarily equal after controlling for other coupons. Table \ref{table:desccoupon} and the tables in Appendix \ref{app:desc} show that the coupons were distributed under consideration of the covariates in the customer registry. We must now assume that the propensity of receiving a coupon (version) differs only depending on observed characteristics, but not on characteristics that are not available to us. This issue can be easily circumvent in practice as long as the information on customers available to marketing campaign planners is also available to those evaluating the campaign.

We note that our assumptions do not rule out inter-temporal spillover effects on customers’ purchasing behavior, since in our main analysis we only examine the (short-term) effect of coupon provision on purchasing behavior during the validity period of the coupon rather than longer-term coupon-induced behavioral shifts. Individuals may, therefore, advance their purchases towards campaign periods in which they receive coupons applicable to the products they are interested in. By including pre-campaign coupon reception and redemption as control variables, we aim at accounting for the fact that previous coupons may influence customer behavior in the outcome period. 

In order to get an impression of the extent to which coupons induce inter-temporal spillover effects and, on the other hand, longer-term increases in customer retention, we also assess the effect of coupon provision in campaign period $t$ on daily expenditures in subsequent periods, namely in $t + 1$ and $t + 2$. It should, however, be noted that the estimated effect is the total effect of coupon reception on purchasing behavior in these subsequent periods, that is, it does not only capture the longer-term coupon-induced change in purchases at the store (net of spillovers from advancing purchases in periods in which the customer has applicable coupons). Rather, it also captures how coupon provision in $t$ affects purchasing behavior in $t + 1$ and $t + 2$ through changing the likelihood of coupon reception in these later periods (e.g., because the customer redeems coupons in $t$ or the coupons incentivize her to increase her purchases in $t$). Disentangling the direct effect of coupon provision on purchasing behavior in subsequent periods from the indirect effect mediated via increasing the likelihood of coupon provision in these later periods would require estimating dynamic treatment effects of treatment sequences, such as the sequence of coupon reception in $t$ and non-reception in $t + 1$ (see \cite*{Bodoryetal2020} for an approach to estimating dynamic treatment effects by means of DML). Further, some coupons valid in $t$ may still be valid in $t + 1$ and even $t + 2$. The estimated effect of coupon provision in $t$ on purchasing behavior in later periods therefore also partially captures the treatment effect of coupons during their validity period. A look at the data shows that the likelihood of having a valid coupon in $t + 1$ or $t + 2$ is highly correlated with that of having a coupon in $t$ (conditional on $X$), be it due to the effect of coupons on re-provision or because the validity period of coupons exceeds that of the artificially created campaign periods. Part of the estimated longer-term effect is therefore likely attributable to the indirect effect of coupon provision in $t$ on daily expenditures in $t + 1$ and $t + 2$, via increasing the probability that the customer has valid coupons in these subsequent periods.


\section{Causal Machine Learning}\label{methodscausalML}

In the following, let $i \in \{1,....,n\}$ be an index for the $n = 1,582 $ customers in the dataset and $t \in \{1,..., T\}$ with $T = 32$ an index for the campaign period. Then, $\{Y_{i,t},D_{i,t},X_{i,t}\}$ denote the outcome, the treatment and the covariates, respectively, for individual $i$ in campaign period $t$. Treatment $D_{i,t}$ is a binary indicator measuring exposure to a coupon campaign (of a specific type) and $Y_{i,t}$ denotes the outcome, defined as average per-day expenditures of customer $i$ in period $t$. The covariates $X_{i,t}$, all measured prior to or at the time of treatment assignment, include socio-economic variables (see Table \ref{table:desccoupon}), the average daily spending by product type in the period prior to the campaign $t-1$, and variables that measure both whether the customer received coupons in $t-1$ and whether he/she redeemed any. For estimating the effect of a particular coupon type, $X_{i,t}$ also contains variables on what other coupon types were provided to the customer in $t$; in addition, it includes information not only about whether the customer received coupons in $t-1$, but also about what type of coupons. 

Under the identifying assumptions outlined in Section \ref{identifyingassumptions}, the ATE $\Delta$ defined in equation \eqref{ATE} corresponds to $\theta$:
\begin{equation}
	\theta = E[\mu(D=1,X) - \mu(D=0,X)] \label{eq:theta}
\end{equation}
where $\mu(D=d,X)$ denotes the conditional mean outcome given treatment state $D=d$ and covariates $X$. As long as the function $\mu$ is of known functional form and $X$ is low-dimensional, we can estimate $\hat{\mu}(D,X)$ by regressing $Y$ on $D$ and $X$ and then determine the ATE according to equation \eqref{eq:theta}. However, the amount of customer data available to marketers is often extensive, and the functional form of relationships between observable customer characteristics and purchasing behavior is often unknown and complex. It may, therefore, in many cases be preferable to use an approach that integrates ML algorithms into the estimation of the causal effect to take advantage of the functional flexibility and the ability to deal with high-dimensional data inherent in  ML algorithms. Put simply, ML algorithms are used to estimate models for predicting  $Y$ as a function of $D$ and $X$ ($\mu(D,X)$) and for predicting the probability of being treated conditional on $X$, which is commonly referred to as the propensity score $p(X) = Pr(D = 1|X)$. These predictions are then integrated into the estimation of the treatment effects.

We assess the causal effect of receiving coupons (of a certain category) on average per-day spending using causal forests, a causal ML method developed by \cite*{WagerAthey2018} that draws on the ML technique of random forests. While the causal forest framework primarily aims at estimating treatment effect heterogeneity, i.e., how the effect of coupons is distributed across different clients and time periods (see Section \ref{methodsheterogeneity}), the estimated causal forests can also be used to estimate the ATE of coupon provision (see Sections \ref{methodsATE}). Both the causal forest algorithm for assessing treatment effect heterogeneity and the estimation procedure used for determining the ATE rely on combining effect estimation on so-called \cite*{Neyman1959}-orthogonal scores with sample splitting. The purpose of orthogonalization is to ensure the robustness of the estimation of causal effects to regularization bias which accrues when using ML to estimate $\mu(D,X)$ and $p(X)$, in the following referred to as plug-in parameters $\eta = (\mu_D(X), p(X))$. Sample splitting, on the other hand, aims to avoid overfitting in the estimation of treatment effects.  In Section \ref{methodsGATE}, we outline how the estimated causal forest can be utilized for determining the treatment effect in different customer segments as defined by selected covariates. Section \ref{methodsoptimalpolicy}, finally, shows how to use the estimated causal forest to determine which customers should optimally be targeted with the different coupon campaigns.

\subsection{Treatment Effect Heterogeneity}\label{methodsheterogeneity}
The causal forest approach  by \cite*{WagerAthey2018} is a modified version of the random forest aimed at determining splitting rules that maximize the heterogeneity of treatment effects in the resulting subsamples. The causal forest provides individualized treatment effect estimates for every observation in the sample as a function of its covariates $X$, which are commonly referred to as Conditional Average Treatment Effects (CATEs), and thereby gives an impression of the heterogeneity in the effect of coupon provision across customers and campaign periods. 

Causal forests are built from so-called causal trees just as random forests are built from regression/classification trees. In order to generate a causal forest, the algorithm repeatedly (2,000 times in our case) draws random samples with 50\% of the observations in the dataset. In each random sample, it estimates a causal tree as follows: first, a randomly selected subset of $min(\sqrt{k} + 20, k)$ covariates is chosen, which in our case amounts for 30 of our $k = 93$ covariates. The algorithm then utilizes these covariates for splitting the sample into two subsamples such that the CATEs in the two resulting subsamples are as heterogeneous as possible. More precisely, the algorithm determines both the covariate and the value  at which the sample should be split (e.g.\ age < 25 vs.\ age $\geq$ 25) to maximize effect heterogeneity. Intuitively, the algorithms considers all possible splits on values of the 30 covariates to find the optimal split in terms of effect heterogeneity. The subsamples obtained from this splitting rule are commonly referred to as nodes. These nodes are further split into a larger number of nodes following the same procedure until some stopping rule is reached, e.g.,\ that no further splits are made if they would entail nodes with less than 5 treated or 5 control observations. The causal forest is finally obtained by averaging over the splitting structure of all 2000 causal trees.

The CATE in the subsamples resulting from each potential split is estimated by means of an approach proposed by \cite*{Robinson1988} that allows estimating the CATE with $\sqrt{n}$ consistency. The approach builds on first predicting the plug-in parameters $\eta = (\mu_D(X), p(X))$, where the plug-in parameters can be estimated using any predictive ML algorithm as long as the plug-in estimates converge with at least a convergence rate of $n^{-1/4}$, and then using the predicted plug-in parameters for estimating the CATE. In our case, the plug-in parameters are predicted by means of regression forests with out-of-bag prediction.\footnote{First, the data set is split into two subsamples, each of which is used to learn regression forests for predicting $\mu_D(X)$ and $p(X)$, respectively. Then, in both subsamples, the plug-in parameters are estimated using the forests learnt in the respective other subsample. The final estimate of the plug-in parameters is obtained by averaging over the estimates from both samples.} In a second step, the algorithm calculates the residuals $Y_{i,t} - \hat{\mu}_{D_{i,t}}(X_{i,t})$ and $D_{i,t} -\hat{p}(X_{i,t})$ for all observations $i,t$ in the random sample used for learning the causal tree. In order to determine the split that maximizes effect heterogeneity in the resulting subsamples, the algorithm regresses $Y_{i,t} - \hat{\mu}_{D_{i,t}}(X_{i,t})$ on $D_{i,t} -\hat{p}(X_{i,t})$ in each subsample. That is, for every potential node, the algorithm estimates the following function, where $\hat{\theta}(X)$ denotes the estimated CATE:\footnote{For computational efficiency, the splitting rules are not determined by estimating the CATEs in all possible subsamples. Rather, the algorithm approximates the between-node effect heterogeneity generated through every potential split by means of a gradient for each observation. Then, the algorithm involves several conditions for formulating splitting rules that aim at avoiding imbalance in the size of the nodes. Explaining these rules in detail would go beyond the scope of this discussion. The manual to the {\tt grf} package, however, provides all the details (see \cite{grfmanual2022}). In our application, we keep all options of the \texttt{causal\_forest} function at their default values.}
\begin{equation}
	Y - \hat{\mu}_{D}(X) = (D -\hat{p}(X)) \hat{\theta}(X).  	\label{eq:residualreg}
\end{equation}
By comparing the estimated CATEs in all potential nodes, the algorithm determines the splitting rule for which the estimated CATEs differ most between the two resulting subsamples. The approach of first predicting the plug-in parameters and then incorporating them into effect estimation ensures that causal effect estimation is more robust to slight approximation errors in the plug-in parameter estimates, which may arise from regularization biases, i.e.,\ from neglecting less important covariates in the splitting procedure.

Furthermore, the causal forest algorithm addresses another source of bias, namely overfitting, i.e., fitting the effect heterogeneity model too strongly to the particularities of the data, such that the procedure picks up not only the actual differences of causal effects across covariates, but also random noise. In order to prevent such overfitting bias, the random sample used for learning a causal tree is itself randomly split into two subsamples, one for building the tree by following the procedure mentioned above, while the other one is used for estimating the treatment effect in every node of the learnt causal tree. That is, by following the splitting rules learnt in the first subsample, the algorithm populates the nodes of the estimated tree with the observations from the second subsample and calculates the CATE in each node based on the observations that fall into the respective node. Trees that are estimated based on this sample splitting procedure are commonly referred to as `honest' trees (because they avoid overfitting).

Through averaging over 2000 causal trees, the causal forest provides the final estimates of the CATEs $\hat{\theta}(X)$, i.e., estimations of individualized treatment effects for every point in $X$. To account for the issue that the behavior of one and the same customer is in general not independent across different campaign periods we cluster standard errors at the customer level. The estimation is performed in the statistical software R (\cite*{R2022}) by means of the {\tt causal\_forest} function provided in the  {\tt grf} package by \cite{grfmanual2022}.

\subsection{Average Treatment Effect}\label{methodsATE}

The estimated causal forest can further be used to identify the ATE of coupon provision and thus to assess the overall effectiveness of the coupons (and that of selected coupon types). \cite*{AtheyWager2019} propose to estimate the ATE by means of a modified version of the Augmented Inverse Probability Weighting (AIPW) estimator, a doubly robust estimator proposed by \cite*{RobinsRotnitzkyZhao95}, that is based on weighting the observations by the inverse of their estimated propensity score. This weighting of observations makes the treatment and the control group comparable in terms of their propensity scores and hence the distribution of relevant covariates $X$ (for more information on the AIPW estimator see e.g. \cite{GlynnQuinn2010}). Double robustness is achieved by estimating the ATE via an orthogonalized function, i.e., the predicted plug-in parameters are included in the estimation such that small estimation errors in either predictor result in an overall negligible error and hence do not introduce bias in the estimation of the ATE. The formula used for estimating the ATE is as follows: 

\begin{eqnarray}
	\hat{\Theta} &=& \dfrac{1}{N  T}\sum_{i \in N, t \in T} \Gamma_{i,t}  \label{eq:orthscore} \\  \notag 	\textrm{ with } \ \Gamma_{i,t} &=& \hat{\theta}(X_{i,t}) + \dfrac{D_{i,t} -  \hat{p}(X_{i,t})}{\hat{p}(X_{i,t})(1 - \hat{p}(X_{i,t}))} \Bigl(Y_{i,t}-\hat{\mu}(X_{i,t}) - \Bigl(D_{i,t} - \hat{p}(X_{i,t})\Bigr)\hat{\theta}(X_{i,t})\Bigr) \notag
\end{eqnarray}

where the plug-in parameters $\hat{\theta}(X), \hat{p}(X)$ and $\hat{\mu}(X)$ for the doubly robust score $ \Gamma_{i,t}$  are obtained from the estimated causal forest. As mentioned above, $\hat{p}(X)$ and $\hat{\mu}(X)$ are predicted by means of regression forests with out-of-bag prediction while  $\hat{\theta}(X)$ is determined using honest trees, i.e., the plug-in estimators for observation $(i,t)$ are computed based on models learnt in samples that do not contain observation $(i,t)$. This makes the AIPW-based ATE estimator robust to regularization bias. Thus, similarly to  how the CATE is estimated for building causal trees, the modified AIPW estimator by \cite*{AtheyWager2019} combines orthogonalization and out-of-sample prediction in order to  address the two sources of bias, overfitting and regularization.





The causal-forest based approach for estimating the ATE described above ensures that the ATE can be estimated with $\sqrt{n}$-consistency, i.e.,\ the estimated ATE converges to the true ATE with a convergence rate of $1/\sqrt{n}$, provided that the ML steps satisfy specific regularity conditions (like $n^{-1/4}$-consistency). A look at equation \eqref{eq:orthscore} reveals that values of $\hat{p}(X_{i,t})$ that are either close to zero or close to one can yield large weights for the respective observations, resulting in unstable performance of the estimator. This issue is commonly addressed by trimming the dataset, i.e., discarding observations with an estimated propensity score that is below or above certain values. A commonly used trimming rule is to remove observations with estimated propensity scores larger than 0.99 or smaller than 0.01, an approach we also employ in this study. 

In our application, we estimate the ATE using the {\tt average\_treatment\_effect} function provided in the  {\tt grf} package for R by \cite{grfmanual2022}, with standard errors clustered at the customer level.

\subsection{Group Average Treatment Effects}\label{methodsGATE}
In order to assess the impact of coupon provision in different customer groups, we also estimate selected Group Average Treatment Effects (GATEs), that is, the average treatment effects in different subgroups as defined by age, income, family size and pre-campaign expenditures, respectively. The variables used to distinguish these subgroups are the age group and family size variables as defined in the original dataset, a variable for average daily expenditures that divides the sample into four subgroups of similar size, and a variable measuring income in broader categories, each of which combines two of the more fine-grained income groups in the original variable. We estimate a linear model of the doubly robust scores $\hat{\Gamma}_{i,t}$ (see equation \eqref{eq:orthscore}) as a function of one of the variables that indicate which subgroup each client belongs to, see \cite{SemenovaChernozhukov2021} for more details.  This approach also allows us to assess effect heterogeneity in customer segments defined by more than one variable, by regressing the \cite*{Neyman1959}-orthogonal scores $\hat{\Gamma}_{i,t}$ on several identifiers (or dummy variables) for belonging to a specific subgroup defined in terms of covariate values (e.g.\ an indicator for being younger female or elderly male customer). We estimate the GATEs by means of the {\tt best\_linear\_projection} function provided in the  {\tt grf}-package.

\subsection{Optimal Policy Learning}\label{methodsoptimalpolicy}
The optimal policy learning approach by \cite*{AtheyWager2021} goes one step further, in the sense that it does not only estimate the effect of coupon provision in predefined customer groups. Rather, it exploits the heterogeneity in coupon effects to determine the coupon distribution rule that maximizes the overall effect of the coupon campaign. Based on observed covariates, the coupon distribution rule distinguishes customer segments that are likely to increase their purchasing behavior upon receiving a coupon from those customer groups not anticipated to respond positively to the campaign. More formally, the algorithm considers specific decision (or policy) rules for whether a coupon should be offered to a customer as a function of the covariate values in $X$, e.g.,\ the customer's age. Let us denote by $\pi(X)$ such a decision rule, which could, for instance, impose that only elderly, but not younger, customers obtain a coupon. 

Mathematically speaking, the rule maps a customer's observed characteristics to the binary treatment decision of whether or not to target the customer through the coupon campaign: $ \pi : X \rightarrow {0, 1}$. Optimal policy learning consists of learning the optimal rule among an assumably limited set of implementable candidate policies, where we use $\Pi$ to denote this set. For instance, another possible rule of how to distribute coupons (in addition to the age-based rule) could be to offer them only to customers with a high volume of previous purchases. Then, both the age- and purchase-dependent rule would enter the set of feasible coupon policies provided in $\Pi$.

For learning the optimal coupon policy, the algorithm of \cite*{AtheyWager2021} use the doubly robust scores $\hat{\Gamma}_{i,t} $ (see equation \eqref{eq:orthscore}). These individual- and time-specific treatment effect estimates are plugged into the following objective function, which aims at maximizing the effectiveness of the coupon campaign by selecting the policy rule with the highest average effect among all policies $\pi$ that are available in the set $\Pi$:
\begin{equation}
	\pi^* = argmax \Bigg\{ \frac{1}{N T} \sum_{i \in {1,...,N}}{}\sum_{t \in {1,...,T}}^{} (2\pi(X_{i,t}) - 1) \hat{\Gamma}_{i,t}: \pi \in \Pi \Bigg\}  \label{eq:objective}
\end{equation}
The optimal policy learning approach does not require defining a priori the policies to be considered, but only the number of customer segments between which coupon allocation can differ and the set of covariates that can be considered for determining these customer segments. Thus, the approach identifies the optimal coupon policy in a data-driven way. To determine the optimal coupon distribution strategy, i.e., the one that maximizes the objective function in \eqref{eq:objective}, the algorithm applies a tree-based approach that considers all possible covariate-defined sample splits for generating the customer segmentation (according to the pre-defined number of segments) and all possible coupon assignment strategies within these segments. The resulting coupon distribution rule can be represented as a decision (or policy) tree, i.e., a tree-shaped graph indicating at which values of which covariate the sample is split and which of the resulting customer segments shall receive coupons.

We estimate decision trees of depth 3, implying that we distinguish 8 customer segments for defining the optimal distribution of coupons by means of the {\tt policytree} package for R by \cite*{Sverdrupetal2020}. For determining the customer segments, we use all the customer characteristics available in the dataset, i.e., age and income group, family size, marital status, and dwelling type. We redefine these variables by setting all missing values to -1, which allows us to omit the variables indicating which observations are missing. Then, we also include the customers' pre-campaign purchasing behavior. Since the algorithm performs a sample split at every possible value of each covariate, i.e., at each observed value, continuous variables can cause performance issues by driving up the number of sample splits. We, therefore, round the pre-campaign average daily expenditures to round values, namely to the nearest 100 for values between 0 and 1,000 and to the nearest 200 for values between 1,000 and 2,000. Further, we group all 157 observations with average daily expenditures of 2,000 or more into one category and include dummies that indicate whether a customer purchased items from the different product categories in the period prior to the campaign. This way, we still capture pre-campaign differences in purchasing behavior well, while substantially reducing the number of sample splits that need to be performed.

\section{Empirical Results}\label{results}
\subsection{Treatment Effect Heterogeneity}\label{resultshetero}
\begin{figure}
	\centering
	\makebox[\textwidth][c]{
		\begin{tabular}{cc}
			\includegraphics[width=.45\textwidth]{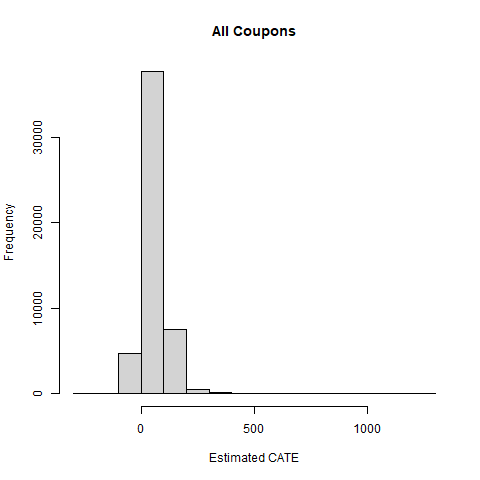}&
			\includegraphics[width=.45\textwidth]{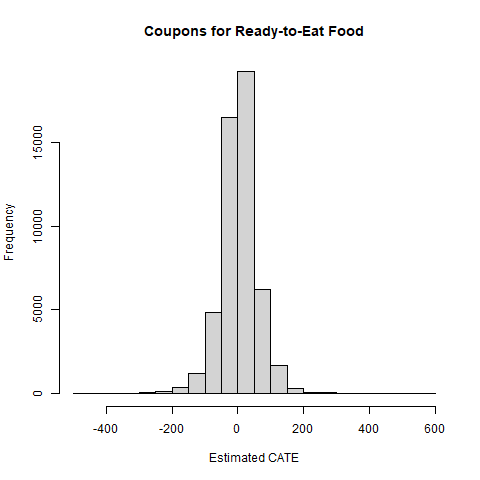} \\[25pt]
	\end{tabular}}
	\makebox[\textwidth][c]{
		\begin{tabular}{cc}
			\includegraphics[width=.45\textwidth]{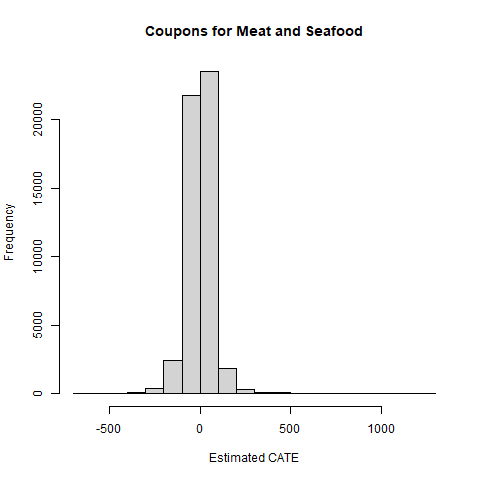}&
			\includegraphics[width=.45\textwidth]{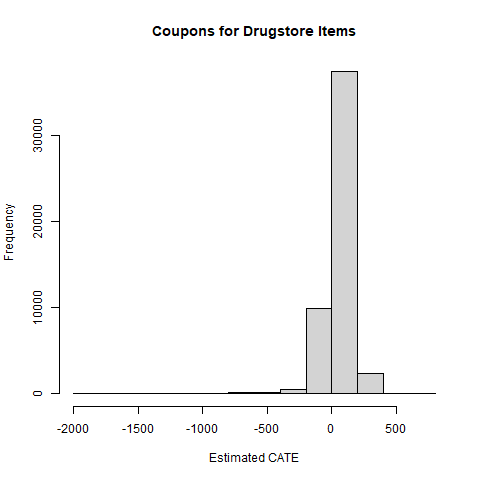} \\[25pt]
	\end{tabular}}
	\makebox[\textwidth][c]{
		\begin{tabular}{cc}
			\includegraphics[width=.45\textwidth]{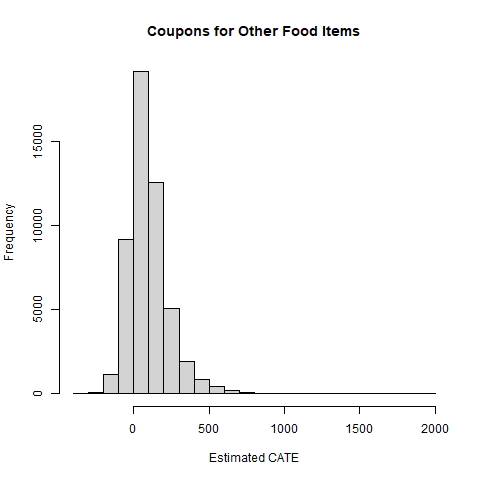}&
			\includegraphics[width=.45\textwidth]{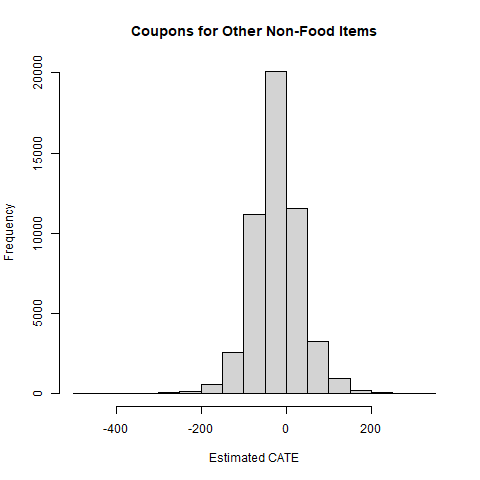} \\[25pt]
	\end{tabular}}
	\caption{Distribution of CATE by coupon type.}
	\label{figure:Histograms}
\end{figure}

Figure \ref{figure:Histograms} shows the distribution of the individualized treatment effects (CATEs) as estimated by means of the causal forest algorithm outlined in Section \ref{methodsheterogeneity}. We can see that the treatment effect of being provided with any coupon is positive for the vast majority of observations and, except for some outliers, ranges between -100 and 200 monetary units. Similarly, provision of drugstore coupons and coupons applicable to other food have a positive effect for the majority of observations. The distribution of coupons applicable to ready-to-eat food as well as meat and seafood, however, seem to be rather centered around zero, with the estimated effect being positive for about half of the observations and negative for the other half. For coupons applicable to other non-food prodcuts, we can even observe a negative effect on daily expenditures for the majority of observations. The plots suggest greater heterogeneity in the treatment effects of the individual coupon categories than when all coupons are analyzed together. It appears that the effects of the different coupon categories cancel each other out to some extent when combined in one analysis, implying that the different coupon categories should best be analyzed separately.

The differences in CATEs as revealed by the causal forest approach suggest not just assessing the ATE, as is done in Section \ref{resultsATE}. Rather, it also invites to analyze how the effect of coupons (of certain categories) differs between  customer groups as defined by covariates $X$ (Section \ref{resultsGATE}) and to learn an optimal coupon distribution scheme that maximizes the expected ATE of coupon provision (Section \ref{resultsPolicytree}).

\subsection{The Causal Effect of Receiving Coupons}\label{resultsATE}
Table \ref{tab:ATE} shows the estimated ATE of receiving any coupon on daily expenditures in the campaign period, as well as that of receiving coupons from each of the five coupon categories, based on the AIPW approach outlined in Section \ref{methodsATE}. The results show that receiving any coupon has a positive and  statistically significant effect on daily expenditures during the campaign period. Providing a customer with a coupon increases her expected daily expenditures by some 63 monetary units. The effect estimates for the different coupon categories provide a more nuanced picture. Provision of coupons for drugstore items and other food has a statistically significant positive effect on daily spending during the campaign period. Receiving coupons that belong to these categories increases expected average daily expenditures during the validity period by some 60 and 75 monetary units, respectively. Handing out coupons applicable to other non-food products, on the other hand, is estimated to decrease a customer's expected average daily expenditures by some 27 monetary units, with this results also being statistically significant. The estimated ATE of providing coupons from the other two categories has no statistically significant effect on the customers' expected daily spending during the campaign period, with the estimated effects being slightly negative.  A possible explanation for the insignificant or significantly negative effect of these latter three coupon types is that the receipt of  such coupons may not incentivize people to buy, but that such coupons are mainly used for products that the coupon recipient would have purchased anyway.

\begin{table}[ht]
	\captionsetup{font=footnotesize}
	\begin{singlespace}
		\centering
		\adjustbox{width = \textwidth, center}{%
			\begin{tabular}{lccc}
				\toprule
				& Coef. & Standard Error & Sign. Level \\ 
				\midrule
				ATE: receiving any coupon & 63.26 & 4.553 & *** \\ 
				\midrule
				ATE: receiving coupon for ready-to-eat food & -2.90 & 8.118 &   \\ 
				ATE: receiving coupon for meat/seafood & -1.42 & 6.045 &   \\ 
				ATE: receiving coupon for other food & 74.74 & 13.559 & *** \\ 
				ATE: receiving coupon for drugstore items & 60.07 & 6.521 & *** \\ 
				ATE: receiving coupon for other non-food items & -26.77 & 6.949 & *** \\ 
				\bottomrule
		\end{tabular}}
		\caption[ATE estimates for coupon reception (general/by coupon type)]{ATE of receiving any coupon as well as the ATEs of receiving coupons applicable to specific product categories, each with standard error and significance level. Significance levels: . p<0.1, * p<0.05, ** p<0.01, *** p<0.001.}\label{tab:ATE}
	\end{singlespace}
\end{table}

As discussed in Section \ref{identifyingassumptions}, coupon provision may, on the one hand, have longer-term positive effects on purchasing behavior by increasing customer loyalty, and on the other hand, bring about inter-temporal spillovers by inducing customers to advance their purchases to periods when they have coupons applicable to them. We therefore also take a look at the overall effect of coupon reception in $t$ on daily expenditures in the following campaign period ($t + 1$) and the period thereafter ($t + 2$) (see Table \ref{tab:ATEpost}). The results suggest that the effect of coupon provision on daily expendidtures is sustainable, i.e., coupon provision in $t$ not only has a short-term effect on purchases in $t$, but also has a statistically significant, albeit smaller, effect on purchases in subsequent periods. This may be due to a coupon-induced increase in customer retention (but also to indirect effects, see the discussion in Section \ref{identifyingassumptions}). 

The longer-term effect of drugstore and other food coupons is also positive and statistically significant, with drugstore coupons showing an even larger effect on purchasing behavior in both post-treatment periods than in the short term. Coupons applicable to other non-food products, that in the short run have a statistically significant negative effect, show a statistically signifificant positive effect on daily spending in the subsequent periods. One possible explanation for this finding is that, while in the short run these coupons were only redeemed for the purchase of products that would have also been purchased without the coupons,  in the longer term they may have increased customer loyalty. 

The estimated effect of meat and seafood coupons on expenditures in $t + 1$ and $ t + 2$ is not statistically significant, while that of ready-to-eat food coupons is even significantly negative for the outcome in $t + 1$, which may indicate spillover effects that are not offset by positive expenditure-increasing effects. For ready-to-eat food and meat/seafood coupons, we can therefore conclude that they do not seem to be an effective marketing tool for increasing customer spending, neither in the short nor in the longer run.

\begin{table}[ht]
	\captionsetup{font=footnotesize}
	\begin{singlespace}
		\centering
		\adjustbox{width = 0.9\textwidth, center}{%
			\begin{tabular}{lcccccc}
				\toprule
				& \multicolumn{3}{c}{Effect in $t + 1$ } & \multicolumn{3}{c}{Effect in  $t + 2$} \\
				& coef. & s.e. & sign. & coef. & s.e. & sign.\\ 
				\midrule
				ATE: receiving any coupon & 39.82 & 3.279 & *** & 34.56 & 3.87 & ***\\ 
				\midrule
				ATE: ready-to-eat food coupons & -28.70 & 6.113 & *** & 4.18 & 8.908 &  \\ 
				ATE: meat/seafood coupons & 6.71 & 6.171 & & 1.13 & 6.244 &   \\ 
				ATE: other food coupons & 52.79 & 11.506 & ***  & 2.46 & 7.345 &   \\ 
				ATE: drugstore coupons & 88.39 & 5.711 & *** & 82.78 & 5.681 & *** \\ 
				ATE: other non-food coupons & 28.03 & 6.204 & *** & 11.64 & 6.017 & . \\  
				\bottomrule
		\end{tabular}}
		\caption[Longer-term ATE estimates]{ATE on daily expenditures in period after coupon campaign ($t + 1$) and the period thereafter ($t + 2$), each with standard error and significance level. Significance levels: . p<0.1, * p<0.05, ** p<0.01, *** p<0.001.}\label{tab:ATEpost}
	\end{singlespace}
\end{table}

In the following, we will again focus on the short-term effect of coupon provision in $t$ on daily expenditures in $t$. The next section examines effect heterogeneity with regard to selected customer characteristics. This is because the provision of coupons could significantly increase spending of certain customer groups, despite not having a statistically significant effect on the overall customer base. Similarly, providing coupons applicable to drugstore or other food could have a significant impact on purchasing behavior only among certain subgroups of customers. 

\FloatBarrier

\subsection{Group Average Treatment Effects}\label{resultsGATE}

\begin{figure}[htp]
	\centering
	\makebox[\textwidth][c]{
		\begin{tabular}{cc}
			\includegraphics[width=.5\textwidth]{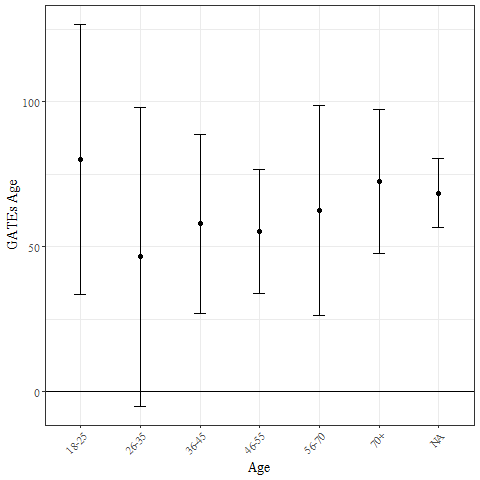}&
			\includegraphics[width=.5\textwidth]{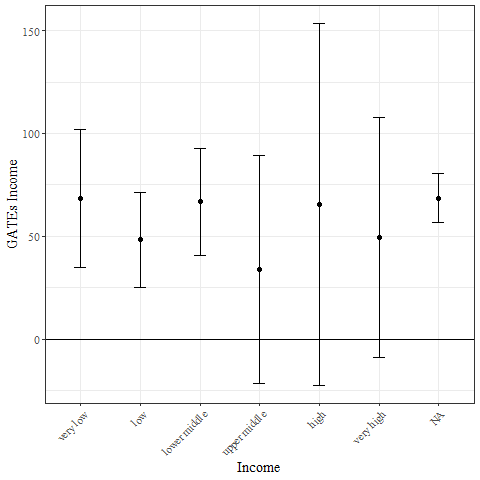} \\[25pt]
	\end{tabular}}
	\makebox[\textwidth][c]{
		\begin{tabular}{cc}
			\includegraphics[width=.5\textwidth]{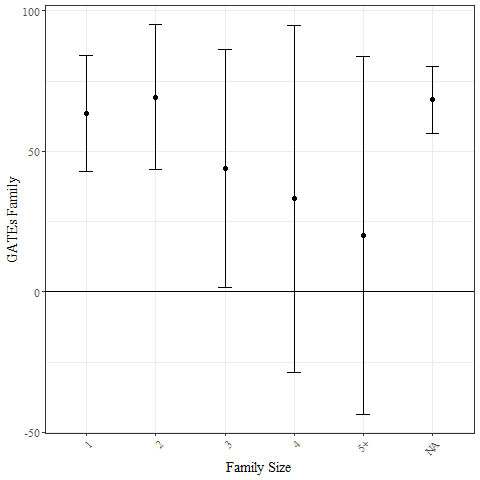}&
			\includegraphics[width=.5\textwidth]{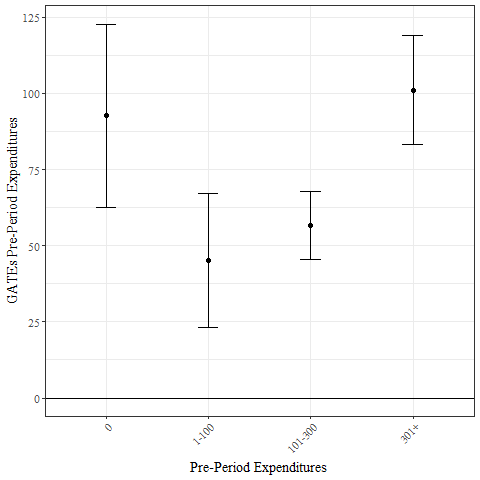} \\[25pt]
	\end{tabular}}
	\caption[GATEs of receiving any coupon]{GATEs of receiving any coupon with 95\% confidence interval.}
	\label{figure:GATE}
\end{figure}

In this section, we assess how the provision of coupons affects different customer groups, based on the approach discussed in Section \ref{methodsGATE}. We illustrate how the effect of providing coupons differs depending on the customers' age, income, family size and pre-campaign expenditures. Further, we also examine the GATEs of those coupon categories with a highly statistically significant ATE, i.e., drugstore coupons and coupons applicable to other food.

Figure \ref{figure:GATE} shows the GATEs of receiving any coupon by age, income, family size and pre-campaign expenditures, respectively. The graphs show that providing coupons has a positive effect on purchasing behavior in every customer group that is statistically significant in most subgroups. The effect of providing coupons tends to be particularly large among customers from smaller households and among those who made either no or large purchases in the period prior to the campaign.

\begin{figure}[htp]
	\centering
	\makebox[\textwidth][c]{
		\begin{tabular}{cc}
			\includegraphics[width=.5\textwidth]{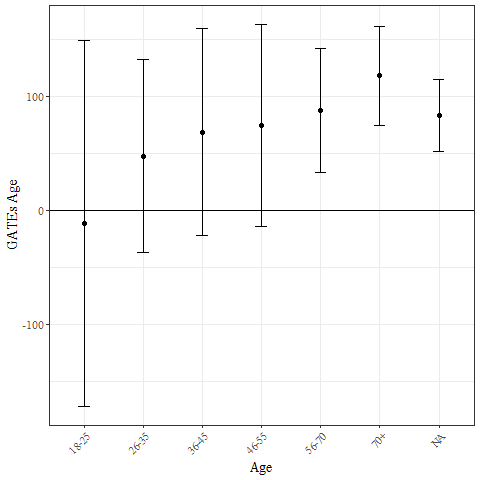}&
			\includegraphics[width=.5\textwidth]{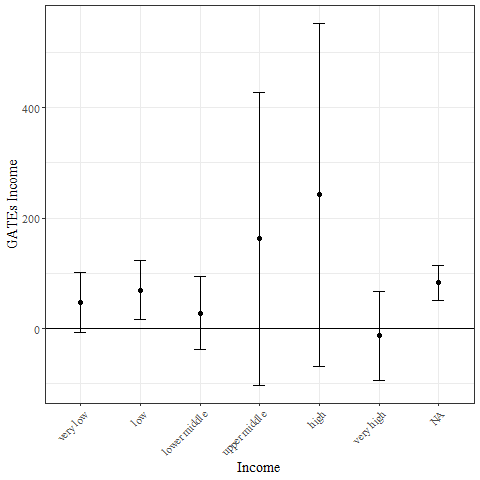} \\[25pt]
	\end{tabular}}
	\makebox[\textwidth][c]{
		\begin{tabular}{cc}
			\includegraphics[width=.5\textwidth]{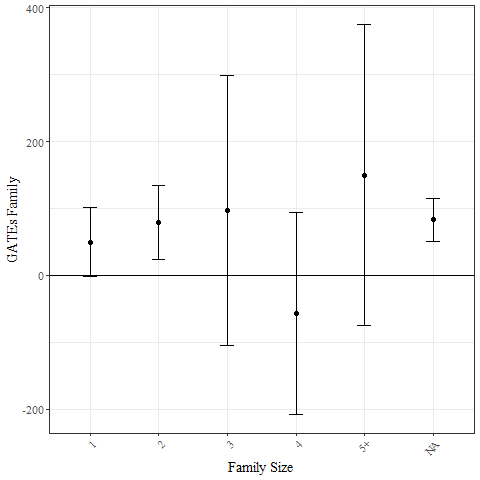}&
			\includegraphics[width=.5\textwidth]{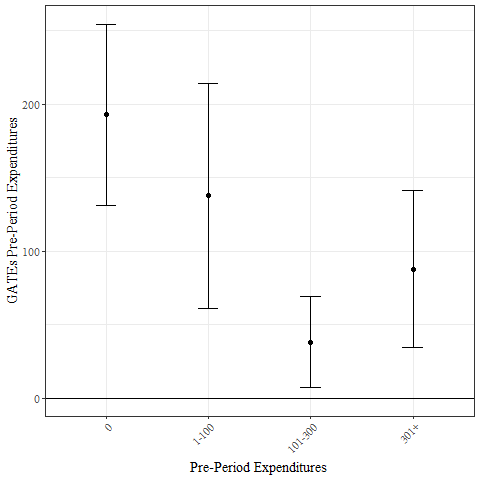} \\[25pt]
	\end{tabular}}
	\caption[GATEs of coupons applicable to other food]{GATEs of coupons applicable to other food items with 95\% confidence interval.}
	\label{figure:GATEOtherFood}
\end{figure}

The GATE charts in Figure \ref{figure:GATEOtherFood} show that in almost all subgroups considered, the provision of coupons for other food has a positive, and in many cases statistically significant, effect on daily spending. The most pronounced differences in GATEs can be found among customer subgroups defined by average daily spending prior to the campaign period. The effect of these food coupons tends to be high and statistically significant for previously inactive customers, while it is much smaller, though still statistically significant, for customers with high pre-period spending. This may suggest that coupons for other food have the potential to reactivate dormant customers. This hypothesis is also supported by the fact that the provision of coupons applicable to other food has a relatively large statistically significant\footnote{the small confidence interval around this GATE estimator can be explained with the large number of observations for which no information on socio-economic characteristics is available} effect among customers for whom information on socio-economic characteristics is not available. Customers for whom no information is available may be more likely to have low loyalty/retention to the store and to be rather inactive customers who can be reactivated by providing them with other food coupons. From these results, we can deduce the hypothesis that coupons applicable to other food are efficient for inactive and non-frequent customers, while they have less impact on the purchasing behavior of frequent shoppers.

\begin{figure}[htp]
	\centering
	\makebox[\textwidth][c]{
		\begin{tabular}{cc}
			\includegraphics[width=.5\textwidth]{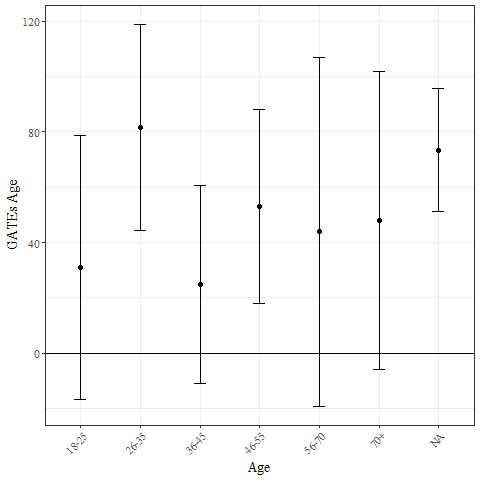}&
			\includegraphics[width=.5\textwidth]{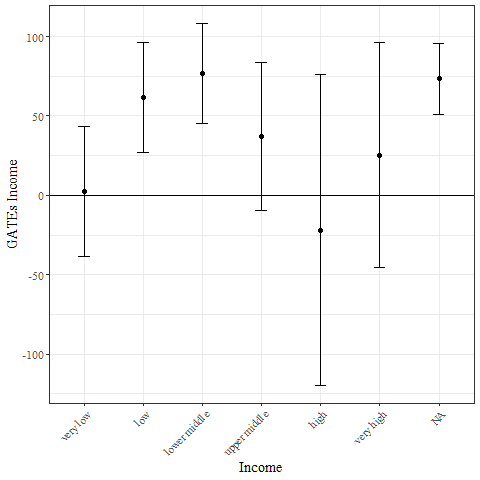} \\[25pt]
	\end{tabular}}
	\makebox[\textwidth][c]{
		\begin{tabular}{cc}
			\includegraphics[width=.5\textwidth]{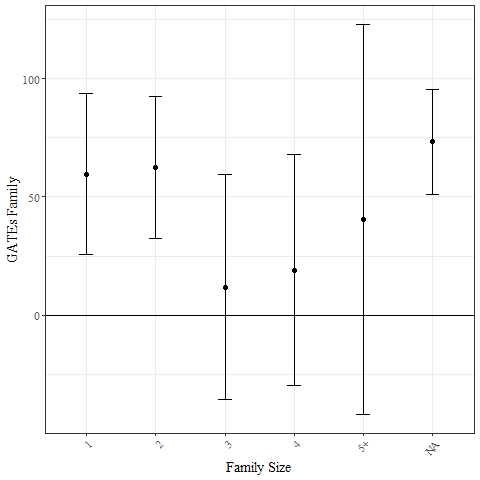}&
			\includegraphics[width=.5\textwidth]{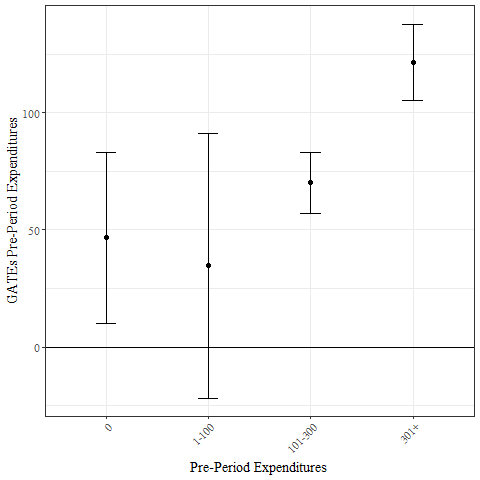} \\[25pt]
	\end{tabular}}
	\caption[GATEs of drugstore coupons]{GATEs of drugstore coupons with 95\% confidence interval.}
	\label{figure:GATEDrugstore}
\end{figure}

Figure \ref{figure:GATEDrugstore} shows that providing drugstore coupons has a positive effect on daily spending for almost all subgroups considered, and that the effect is statistically significant in most cases. Again, the largest difference can be found in the GATE estimates by pre-campaign spending. The effect of drugstore coupons on average per-day expenditures is larger the higher the customer's pre-campaign spending, which is the reverse pattern of what we find for coupons applicable to other food. This suggests that other food coupons are more efficient at reactivating dormant customers and drugstore coupons at retaining frequent shoppers. The GATE plots for the other three coupon categories can be found in Appendix \ref{appendix:GATE}.

While  ML-based estimation of ATEs and GATEs is an excellent tool for evaluating the effect of coupon campaigns, it is not necessarily most appropriate for deriving strategies for later coupon campaigns. For this purpose, the optimal policy learning framework by \cite*{AtheyWager2021} is arguably superior as it determines which customer groups to provide with coupons in order to maximize the ATE.

\subsection{The Optimal Distribution of Coupons}\label{resultsPolicytree}

\begin{figure}[htp]
	\centering
	\makebox[\textwidth][c]{
		\begin{tabular}{cc}
			\includegraphics[width=.57\textwidth]{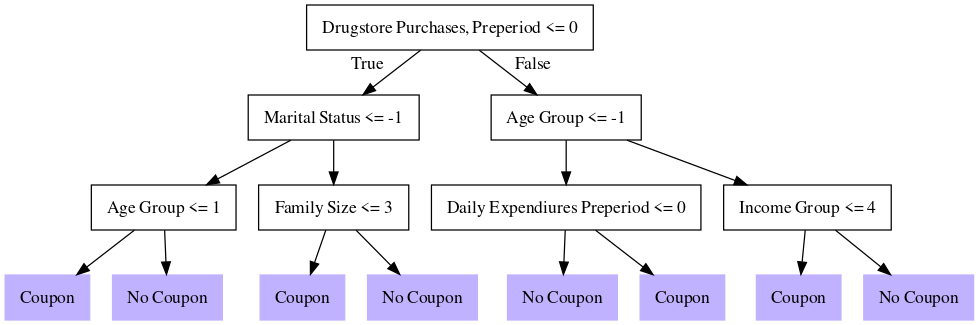}&
			\includegraphics[width=.63\textwidth]{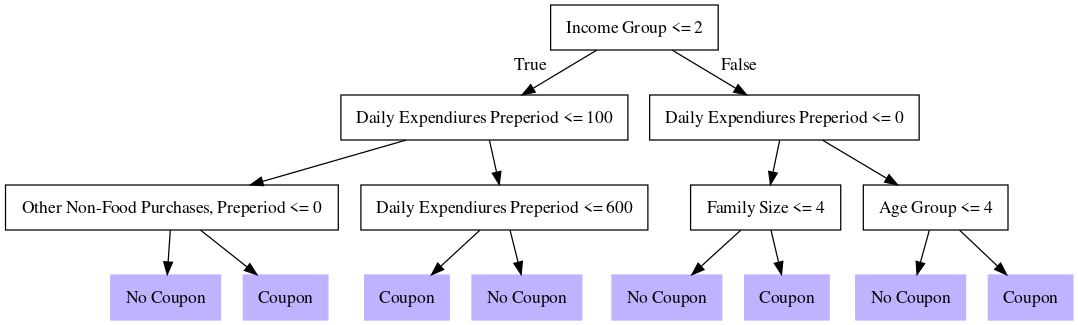} \\
			\textbf{(a)}  & \textbf{(b)} \\[25pt]
	\end{tabular}}
	\makebox[\textwidth][c]{
		\begin{tabular}{cc}
			\includegraphics[width=.6\textwidth]{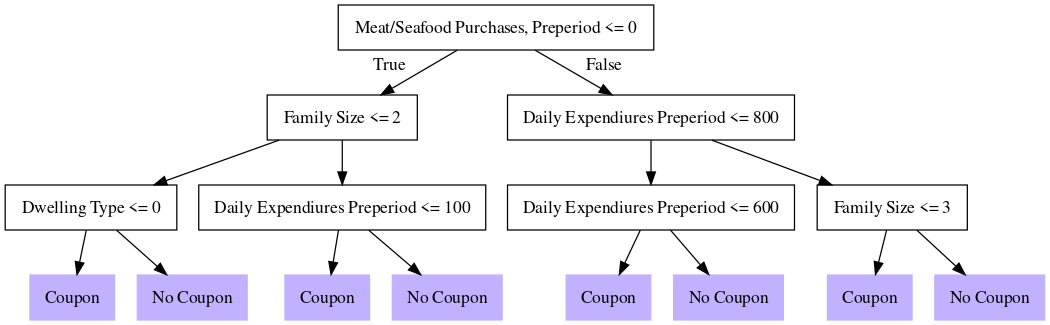}&
			\includegraphics[width=.60\textwidth]{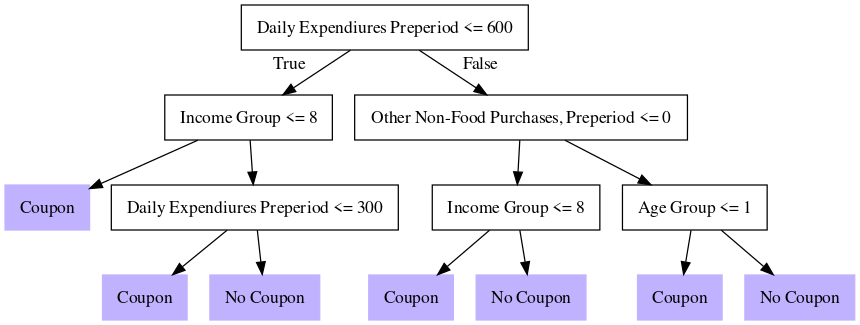} \\
			\textbf{(c)}  & \textbf{(d)} \\[25pt]
	\end{tabular}}
	\begin{tabular}{ccc}
		&\includegraphics[width=.65\textwidth]{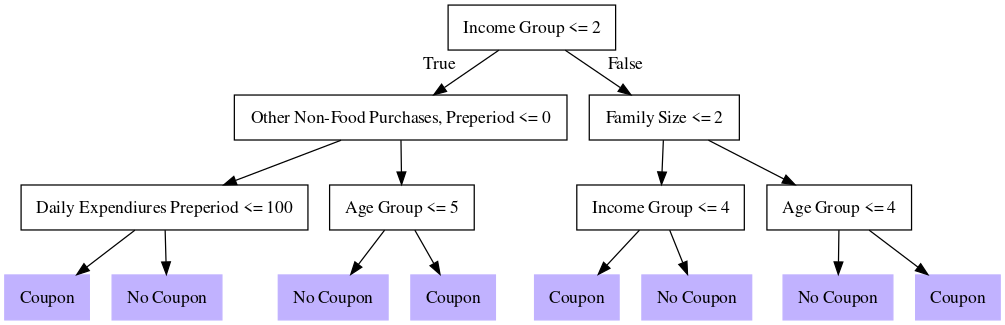}&\\
		&\textbf{(e)}&\\
	\end{tabular}
	\caption[Depth-3 trees for coupon provision, by coupon type]{Depth-3 trees for coupons applicable to (a) ready-to-eat food, (b) meat and seafood, (c) other food, (d) drugstore products as well as (e) other non-food products.}
	\label{figure:trees}
\end{figure}

Figure \ref{figure:trees} shows the optimal distribution rules (or policies) for each coupon category as identified based on the optimal policy learning approach outlined in Section \ref{methodsoptimalpolicy}. The optimal distribution rule for ready-to-eat food coupons (decision tree (a)) suggests providing ready-to-eat food coupons to customers with no drugstore purchases in the pre-campaign period if their marital status is unknown\footnote{Please note that for family size, marital status, age group and income group the value -1 denotes that information about these variables is unavailable, see also the description of the methodology in Section \ref{methodsoptimalpolicy}} and their age is unknown or they are not older than 26, or if their marital status is known and they live in a household of no more than three members. The retailer should further provide ready-to-eat food coupons to customers who purchased drugstore products in the pre-period if their income is in one of the lowest four income groups or unknown, or if their age is unknown and their average daily purchases in the pre-period were less than 50 monetary units\footnote{Please note that daily expenditures are rounded for creating policy trees. All customers with daily expenditures below 50 monetary units fulfil the condition `Daily Expenditure Preperiod <= 0'}.

The optimal distribution rule for drugstore coupons (decision tree (d)) proposes providing drugstore coupons to those customers with unkown, low, or middle incomes if their daily pre-campaign expenditures did not exceed 600 monetary units. Customers belonging to the high-income group should receive drugstore coupons if their average in-store spending did not exceed 300 monetary units per day in the period before the campaign. In addition, customers whose pre-campaign expenditures exceeded 600 monetary units per day should only be provided with drugstore coupons if they did not purchase any other non-food products in the pre-campaign period and do not belong to the high-income group, or if they purchased non-food products and are either 18-26 years old or of unknown age.

The distribution rules paint a similar picture as the GATE estimates in Section \ref{resultshetero} about which customer groups are likely to be positively impacted by the provision of certain coupon types. In contrast to the assessment of effect heterogeneity across pre-specified broad categories in Section \ref{resultshetero}, the optimal policy learning algorithm finds the covariate values at which the sample should optimally be split in order to maximize the ATE and defines groups of coupon recipients and non-recipients based on multiple covariates.

The other decision trees can be interpreted accordingly. A look at the covariates used for sample splitting in those other decision trees shows that each observed customer characteristic is used for distribution rules of at least one coupon type.

\subsection{Robustness Checks}\label{resultsrobustness}
As described in Section \ref{data}, our dataset contains a large number of observations with missing socio-economic information. To investigate the robustness of our results with respect to these missing values, we performed the entire analysis on a reduced dataset containing only observations of customers whose socio-economic background is known, i.e., on a dataset with 13,792 observations of the purchasing behavior of $ n = 431$ individuals. The estimated ATEs can be found in Table \ref{tab:robustnessATE}. They are close to the ATE estimates from the full dataset, although the standard errors are of course considerably larger - due to the much smaller number of observations.
\begin{table}[ht]
	\captionsetup{font=footnotesize}
	\begin{singlespace}
		\centering
		\adjustbox{width = 0.9\textwidth, center}{%
			\begin{tabular}{llll}
				\toprule
				& Coef. & Standard Error & Sign. Level \\ 
				\midrule
				ATE: receiving any coupon & 41.63 & 8.024 & *** \\ 
				\midrule
				ATE: receiving coupon for ready-to-eat food & -6.50 & 10.618 &   \\ 
				ATE: receiving coupon for meat and seafood & -26.94 & 9.374 & ** \\ 
				ATE: receiving coupon for other food & 96.72 & 22.897 & *** \\ 
				ATE: receiving coupon for drugstore items & 41.12 & 9.493 & *** \\ 
				ATE: receiving coupon for other non-food items & -22.72 & 9.568 & * \\ 
				\bottomrule
		\end{tabular}}
		\caption[ATE of coupon reception (general/by coupon type), estimated in reduced dataset]{ATE of receiving any coupon as well as the ATEs of receiving coupons applicable to specific product categories in the reduced dataset (without observations with missing socio-economic information), each with standard error and significance level. Significance levels: . p<0.1, * p<0.05, ** p<0.01, *** p<0.001.}\label{tab:robustnessATE}
	\end{singlespace}
\end{table}

\begin{table}[ht]
	\captionsetup{font=footnotesize}
	\begin{singlespace}
		\centering
		\adjustbox{width = 0.9\textwidth, center}{%
			\begin{tabular}{lcccccc}
				\toprule
				& \multicolumn{3}{c}{Effect in $t + 1$ } & \multicolumn{3}{c}{Effect in  $t + 2$} \\
				& coef. & s.e. & sign. & coef. & s.e. & sign.\\ 
				\midrule
				ATE: receiving any coupon & 24.29 & 7.609 & ** & 20.20 & 8.545 & * \\ 
				\midrule
				ATE: ready-to-eat food coupons & -19.31 & 9.878 & . & -26.22 & 9.586 & ** \\ 
				ATE: meat/seafood coupons & -7.90 & 10.305 &   & -8.56 & 11.434 &   \\ 
				ATE: other food coupons & 54.77 & 14.86 & ***  & -4.28 & 15.827 &  \\ 
				ATE: drugstore coupons & 68.69 & 8.398 & *** & 47.31 & 11.723 & ***  \\ 
				ATE: other non-food coupons & 10.33 & 7.235 & & 6.96 & 10.275 &    \\ 
				\bottomrule
		\end{tabular}}
		\caption[Longer-term ATE estimates, estimated in reduced dataset]{ATE on daily expenditures in period after each coupon campaign ($t + 1$) and the period thereafter ($t + 2$), estimated in the reduced dataset (without observations with missing socio-economic information), each with standard error and significance level. Significance levels: . p<0.1, * p<0.05, ** p<0.01, *** p<0.001.}\label{tab:robustnessATEpost}
	\end{singlespace}
\end{table}

The GATE and policy tree plots are provided in Appendix \ref{appendix:Robustness}. The GATE plots show similar patterns in how different customer groups are affected by each coupon type, although of course they are not exactly identical with the GATEs estimated in the full sample. The policy tree plots show that the splitting rules are based on a similar set of variables with similar cutting points as those estimated in the full dataset. These results suggest that the large number of observations with missing socio-economic information does not introduce systematic bias into the estimation of the treatment effects and the optimal coupon distribution scheme.

\section{Conclusion}\label{conclusion}
This paper presented different causal ML methods with multiple application possibilities in marketing research and business development. The application of these methods to evaluate a retailer's coupon campaign and optimize the distribution of coupons illustrated their potential in identifying the ATE of coupon provision, effect heterogeneity, as well as optimal coupon distribution rules. 

We could, for instance, find that only coupons belonging to two out of five categories, namely those applicable to  drugstore and other food, have a positive and statistically significant overall effect on purchases, while receiving coupons for other non-food products actually significantly reduces customers' daily spending. Additionally, we were able to pinpoint different customer subgroups whose purchasing behavior can be influenced particularly strongly through provision of certain types of coupons and who should therefore be optimally addressed with the corresponding coupon campaigns. This information would enable the retailer to optimally target her coupon campaigns, i.e., such that the overall effect is maximized. 

The proposed causal ML methods can further be applied to evaluate and optimize a variety of other marketing and business strategies, requiring only observational data from the context of previous campaigns or business decisions, and utilizing all available (Big) data, whether structured or unstructured. Other potential applications for the proposed causal ML methods are the evaluation and optimization of targetable (online) marketing campaigns, loyalty programs and campaigns for dealing with customer attrition, but also the assessment of different employee benefit plans, designs of job postings, or in-house training programs. The potential use cases of causal ML in business and marketing are manifold.
\pagebreak

\bibliographystyle{econometrica}
\bibliography{Literature}

@article{Rubin1974,
  title={Estimating causal effects of treatments in randomized and nonrandomized studies.},
  author={Rubin, Donald B},
  journal={Journal of educational Psychology},
  volume={66},
  number={5},
  pages={688},
  year={1974},
  publisher={American Psychological Association}
}

@ARTICLE{RobinsRotnitzkyZhao95,
	author    = "J. Robins and A. Rotnitzky and L. Zhao",
	title	  = "Analysis of Semiparametric Regression Models for Repeated Outcomes in the Presence of Missing Data",
	year      = 1995,
	journal   = "Journal of American Statistical Association",
	volume    = 90,
	pages     = "106-121"}

@article{Bodoryetal2020,
	title={Evaluating (weighted) dynamic treatment effects by double machine learning},
	author={Bodory, Hugo and Huber, Martin and Laff{\'e}rs, Luk{\'a}{\v{s}}},
	journal={arXiv preprint arXiv:2012.00370},
	year={2020}
}

@article{Sobel2006,
	author = {Michael E Sobel},
	title = {What Do Randomized Studies of Housing Mobility Demonstrate?},
	journal = {Journal of the American Statistical Association},
	volume = {101},
	pages = {1398-1407},
	year  = {2006},
}

@article{HudgensHalloran2008,
	author = {Michael G Hudgens and M. Elizabeth Halloran},
	title = {Toward Causal Inference With Interference},
	journal = {Journal of the American Statistical Association},
	volume = {103},
	pages = {832-842},
	year  = {2008},
}

@article{HongRaudenbush2006,
	author = {Guanglei Hong and Stephen W Raudenbush},
	title = {Evaluating Kindergarten Retention Policy},
	journal = {Journal of the American Statistical Association},
	volume = {101},
	pages = {901-910},
	year  = {2006},
}

@article{Im00,
  AUTHOR =       {G. W. Imbens},
  TITLE =        {The role of the propensity score in estimating dose-response functions},
  JOURNAL =      {Biometrika},
  YEAR =         {2000},
  volume =       {87},
  pages =        {706-710},
}

@inproceedings{Le01,
  AUTHOR =       {M. Lechner},
  TITLE =        {Identification and estimation of causal effects of multiple treatments under the conditional independence assumption},
  BOOKTITLE =    {Econometric Evaluations of Active Labor Market Policies in Europe},
  YEAR =         {2001},
  editor =       {M Lechner and F Pfeiffer},
  publisher =    {Heidelberg: Physica},
}

@article{RosenbaumRubin1983,
	title={Assessing sensitivity to an unobserved binary covariate in an observational study with binary outcome},
	author={Rosenbaum, Paul R and Rubin, Donald B},
	journal={Journal of the Royal Statistical Society: Series B (Methodological)},
	volume={45},
	number={2},
	pages={212--218},
	year={1983},
	publisher={Wiley Online Library}
}

@article{Neyman23,
	author    = {Neyman, Jerzy},
	title	  = {On the Application of Probability Theory to Agricultural Experiments. Essay on Principles.},
	year      = {1923},
	journal   = {Statistical Science},
	volume    = {Reprint, 5},
	pages     = {463-480}
}

@article{WagerAthey2018,
	author = {Stefan Wager and Susan Athey},
	title = {Estimation and Inference of Heterogeneous Treatment Effects using Random Forests},
	journal = {Journal of the American Statistical Association},
	volume = {113},
	pages = {1228-1242},
	year  = {2018},
}

@ARTICLE{AtheyTibshiraniWager2019,
  author =       {Susan Athey and Julie Tibshirani and Stefan Wager},
  title =        {Generalized random forests},
  journal =      {The Annals of Statistics},
  year =         {2019},
  volume =       {47},
  pages =        {1148-1178},
}

@article{VanderWeeleetal2013,
	title={Causal inference under multiple versions of treatment},
	author={VanderWeele, Tyler J and Hernan, Miguel A},
	journal={Journal of causal inference},
	volume={1},
	number={1},
	pages={1--20},
	year={2013},
	publisher={De Gruyter}
}

@article{Rubin1980,
  title={Randomization analysis of experimental data: The Fisher randomization test comment},
  author={Rubin, Donald B},
  journal={Journal of the American Statistical Association},
  volume={75},
  number={371},
  pages={591--593},
  year={1980},
  publisher={JSTOR}
}

@article{Andrewsetal2016,
	title={Mobile ad effectiveness: Hyper-contextual targeting with crowdedness},
	author={Andrews, Michelle and Luo, Xueming and Fang, Zheng and Ghose, Anindya},
	journal={Marketing Science},
	volume={35},
	number={2},
	pages={218--233},
	year={2016},
	publisher={INFORMS}
}

@inproceedings{HeJiang2017,
	title={Understanding Users' Coupon Usage Behaviors in E-Commerce Environments},
	author={He, Jiawei and Jiang, Wenjun},
	booktitle={2017 IEEE International Symposium on Parallel and Distributed Processing with Applications and 2017 IEEE International Conference on Ubiquitous Computing and Communications (ISPA/IUCC)},
	pages={1047--1053},
	year={2017},
	organization={IEEE}
}

@article{Donnellyetal2021,
	title={Counterfactual inference for consumer choice across many product categories},
	author={Donnelly, Robert and Ruiz, Francisco JR and Blei, David and Athey, Susan},
	journal={Quantitative Marketing and Economics},
	pages={1--39},
	year={2021},
	publisher={Springer}
}

@article{Xiaetal2019,
	title={Using conditional restricted Boltzmann machines to model complex consumer shopping patterns},
	author={Xia, Feihong and Chatterjee, Rabikar and May, Jerrold H},
	journal={Marketing Science},
	volume={38},
	number={4},
	pages={711--727},
	year={2019},
	publisher={INFORMS}
}

@article{Huetal2019,
	title={Search and learning at a daily deals website},
	author={Hu, Mantian and Dang, Chu and Chintagunta, Pradeep K},
	journal={Marketing Science},
	volume={38},
	number={4},
	pages={609--642},
	year={2019},
	publisher={INFORMS}
}

@article{Arevalillo2021,
	title={Ensemble learning from model based trees with application to differential price sensitivity assessment},
	author={Arevalillo, Jorge M},
	journal={Information Sciences},
	volume={557},
	pages={16--33},
	year={2021},
	publisher={Elsevier}
}

@article{Ramzanetal2019,
	title={An intelligent data analysis for recommendation systems using machine learning},
	author={Ramzan, Bushra and Bajwa, Imran Sarwar and Jamil, Noreen and Amin, Riaz Ul and Ramzan, Shabana and Mirza, Farhan and Sarwar, Nadeem},
	journal={Scientific Programming},
	volume={2019},
	year={2019},
	publisher={Hindawi}
}

@article{AnithaKalaiarasu2021,
	title={Optimized machine learning based collaborative filtering (OMLCF) recommendation system in e-commerce},
	author={Anitha, J and Kalaiarasu, M},
	journal={Journal of Ambient Intelligence and Humanized Computing},
	volume={12},
	number={6},
	pages={6387--6398},
	year={2021},
	publisher={Springer}
}

@article{GordiniVeglio2017,
	title = {Customers churn prediction and marketing retention strategies. An application of support vector machines based on the AUC parameter-selection technique in B2B e-commerce industry},
	journal = {Industrial Marketing Management},
	volume = {62},
	pages = {100-107},
	year = {2017},
	issn = {0019-8501},
	doi = {https://doi.org/10.1016/j.indmarman.2016.08.003},
	url = {https://www.sciencedirect.com/science/article/pii/S0019850116301651},
	author = {Niccolò Gordini and Valerio Veglio}}

@inproceedings{Luketal2019,
	title={Design of an intelligent customer identification model in e-Commerce logistics industry},
	author={Luk, CC and Choy, KL and Lam, HY},
	booktitle={MATEC Web of Conferences},
	volume={255},
	pages={04003},
	year={2019},
	organization={EDP Sciences}
}

@article{Greensteinetal2017,
	title={Personal-discount sensitivity prediction for mobile coupon conversion optimization},
	author={Greenstein-Messica, Asnat and Rokach, Lior and Shabtai, Asaf},
	journal={Journal of the Association for Information Science and Technology},
	volume={68},
	number={8},
	pages={1940--1952},
	year={2017},
	publisher={Wiley Online Library}
}

@article{InmanMcAlister1994,
	title={Do coupon expiration dates affect consumer behavior?},
	author={Inman, J Jeffrey and McAlister, Leigh},
	journal={Journal of Marketing Research},
	volume={31},
	number={3},
	pages={423--428},
	year={1994},
	publisher={SAGE Publications Sage CA: Los Angeles, CA}
}

@article{Krishnaetl1999,
	title={Short-or long-duration coupons: The effect of the expiration date on the profitability of coupon promotions},
	author={Krishna, Aradhna and Zhang, Z John},
	journal={Management Science},
	volume={45},
	number={8},
	pages={1041--1056},
	year={1999},
	publisher={INFORMS}
}

@article{Rajuetal1994,
	title={The effect of package coupons on brand choice},
	author={Raju, Jagmohan S and Dhar, Sanjay K and Morrison, Donald G},
	journal={Marketing Science},
	volume={13},
	number={2},
	pages={145--164},
	year={1994},
	publisher={INFORMS}
}

@article{LeoneSrinivasan1996,
	title={Coupon face value: Its impact on coupon redemptions, brand sales, and brand profitability},
	author={Leone, Robert P and Srinivasan, Srini S},
	journal={Journal of retailing},
	volume={72},
	number={3},
	pages={273--289},
	year={1996},
	publisher={Elsevier}
}

@article{Marianietal2021,
	title={AI in marketing, consumer research and psychology: A systematic literature review and research agenda},
	author={Mariani, Marcello M and Perez-Vega, Rodrigo and Wirtz, Jochen},
	journal={Psychology \& Marketing},
	year={2021},
	publisher={Wiley Online Library}
}

@article{MaSun2020,
	title={Machine learning and AI in marketing--Connecting computing power to human insights},
	author={Ma, Liye and Sun, Baohong},
	journal={International Journal of Research in Marketing},
	volume={37},
	number={3},
	pages={481--504},
	year={2020},
	publisher={Elsevier}
}

@article{HairSarstedt2021,
	title={Data, measurement, and causal inferences in machine learning: opportunities and challenges for marketing},
	author={Hair Jr, Joseph F and Sarstedt, Marko},
	journal={Journal of Marketing Theory and Practice},
	volume={29},
	number={1},
	pages={65--77},
	year={2021},
	publisher={Taylor \& Francis}
}

@article{Mustaketal2021,
	title={Artificial intelligence in marketing: Topic modeling, scientometric analysis, and research agenda},
	author={Mustak, Mekhail and Salminen, Joni and Pl{\'e}, Lo{\"\i}c and Wirtz, Jochen},
	journal={Journal of Business Research},
	volume={124},
	pages={389--404},
	year={2021},
	publisher={Elsevier}
}

@article{RubinRichard2006,
	title={Estimating the causal effects of marketing interventions using propensity score methodology},
	author={Rubin, Donald B and Waterman, Richard P},
	journal={Statistical Science},
	pages={206--222},
	year={2006},
	publisher={JSTOR}
}

@article{Biswasetal2013,
	title={Consumer evaluations of sale prices: role of the subtraction principle},
	author={Biswas, Abhijit and Bhowmick, Sandeep and Guha, Abhijit and Grewal, Dhruv},
	journal={Journal of Marketing},
	volume={77},
	number={4},
	pages={49--66},
	year={2013},
	publisher={SAGE Publications Sage CA: Los Angeles, CA}
}

@article{Zhengetal2021,
	title={Retail price discount depth and perceived quality uncertainty},
	author={Zheng, Dan and Chen, Yuxin and Zhang, Zhe and Che, Hai},
	journal={Journal of Retailing},
	year={2021},
	publisher={Elsevier}
}

@article{ChoiCoulter2012,
	title={It's not all relative: the effects of mental and physical positioning of comparative prices on absolute versus relative discount assessment},
	author={Choi, Pilsik and Coulter, Keith S},
	journal={Journal of Retailing},
	volume={88},
	number={4},
	pages={512--527},
	year={2012},
	publisher={Elsevier}
}

@article{AndersonSimester2004,
	title={Long-run effects of promotion depth on new versus established customers: three field studies},
	author={Anderson, Eric T and Simester, Duncan I},
	journal={Marketing Science},
	volume={23},
	number={1},
	pages={4--20},
	year={2004},
	publisher={INFORMS}
}

@article{Jiaetal2018,
	title={Do consumers always spend more when coupon face value is larger? The inverted U-shaped effect of coupon face value on consumer spending level},
	author={Jia, He and Yang, Sha and Lu, Xianghua and Park, C Whan},
	journal={Journal of Marketing},
	volume={82},
	number={4},
	pages={70--85},
	year={2018},
	publisher={SAGE Publications Sage CA: Los Angeles, CA}
}

@article{GopalakrishnanPark2021,
	title={The impact of coupons on the visit-to-purchase funnel},
	author={Gopalakrishnan, Arun and Park, Young-Hoon},
	journal={Marketing Science},
	volume={40},
	number={1},
	pages={48--61},
	year={2021},
	publisher={INFORMS}
}

@article{Spiekermannetal2011,
	title={Street marketing: how proximity and context drive coupon redemption},
	author={Spiekermann, Sarah and Rothensee, Matthias and Klafft, Michael},
	journal={Journal of Consumer Marketing},
	year={2011},
	publisher={Emerald Group Publishing Limited}
}

@article{ReimersXie2019,
	title={Do coupons expand or cannibalize revenue? Evidence from an e-Market},
	author={Reimers, Imke and Xie, Claire},
	journal={Management Science},
	volume={65},
	number={1},
	pages={286--300},
	year={2019},
	publisher={INFORMS}
}

@article{Halvorsenetal2016,
	title={Reducing subway crowding: analysis of an off-peak discount experiment in Hong Kong},
	author={Halvorsen, Anne and Koutsopoulos, Haris N and Lau, Stephen and Au, Tom and Zhao, Jinhua},
	journal={Transportation Research Record},
	volume={2544},
	number={1},
	pages={38--46},
	year={2016},
	publisher={SAGE Publications Sage CA: Los Angeles, CA}
}

@article{Huberetal2021,
	title={Business analytics meets artificial intelligence: Assessing the demand effects of discounts on Swiss train tickets},
	author={Huber, Martin and Meier, Jonas and Wallimann, Hannes},
	journal={arXiv preprint arXiv:2105.01426},
	year={2021}
}

@techreport{Zhangetal2017,
	title={How does dynamic pricing affect customer behavior on retailing platforms? evidence from a large randomized experiment on alibaba},
	author={Zhang, Dennis J and Dai, Hengchen and Dong, Lingxiu and Qi, Fangfang and Zhang, Nannan and Liu, Xiaofei and Liu, Zhongyi},
	year={2017},
	institution={Working paper, SSRN}
}

@article{PusztovaBabic2020,
	title={Performance Assessment of Different Classification Methods for Coupon Marketing in E-Commerce},
	author={Pusztov{\'a}, L'udmila and Babi{\v{c}}, Franti{\v{s}}ek},
	journal={Acta Electrotechnica et Informatica},
	volume={20},
	number={3},
	pages={11--16},
	year={2020}
}

@article{Renetal2021,
	title={A two-stage model for forecasting consumers’ intention to purchase with e-coupons},
	author={Ren, Xinxin and Cao, Jingjing and Xu, Xianhao and others},
	journal={Journal of Retailing and Consumer Services},
	volume={59},
	pages={102289},
	year={2021},
	publisher={Elsevier}
}

@article{Koehnetal2020,
	title={Predicting online shopping behaviour from clickstream data using deep learning},
	author={Koehn, Dennis and Lessmann, Stefan and Schaal, Markus},
	journal={Expert Systems with Applications},
	volume={150},
	pages={113342},
	year={2020},
	publisher={Elsevier}
}

@inproceedings{Xiaoetal2021,
	title={DMBGN: Deep Multi-Behavior Graph Networks for Voucher Redemption Rate Prediction},
	author={Xiao, Fengtong and Li, Lin and Xu, Weinan and Zhao, Jingyu and Yang, Xiaofeng and Lang, Jun and Wang, Hao},
	booktitle={Proceedings of the 27th ACM SIGKDD Conference on Knowledge Discovery \& Data Mining},
	pages={3786--3794},
	year={2021}
}

@article{Danaheretal2015,
	title={Where, when, and how long: Factors that influence the redemption of mobile phone coupons},
	author={Danaher, Peter J and Smith, Michael S and Ranasinghe, Kulan and Danaher, Tracey S},
	journal={Journal of Marketing Research},
	volume={52},
	number={5},
	pages={710--725},
	year={2015},
	publisher={SAGE Publications Sage CA: Los Angeles, CA}
}

@article{Neyman1959,
	title={Optimal asymptotic tests of composite hypotheses},
	author={Neyman, Jerzy},
	journal={Probability and statsitics},
	pages={213--234},
	year={1959},
	publisher={Wiley}
}

@article{Anderson2008,
	title={The end of theory: The data deluge makes the scientific method obsolete},
	author={Anderson, Chris},
	journal={Wired magazine},
	volume={16},
	number={7},
	pages={16--07},
	year={2008}
}

@article{Erevellesetal2016,
	title={Big Data consumer analytics and the transformation of marketing},
	author={Erevelles, Sunil and Fukawa, Nobuyuki and Swayne, Linda},
	journal={Journal of business research},
	volume={69},
	number={2},
	pages={897--904},
	year={2016},
	publisher={Elsevier}
}

@misc{Lycett2013,
	title={‘Datafication’: making sense of (big) data in a complex world},
	author={Lycett, Mark},
	journal={European Journal of Information Systems},
	volume={22},
	number={4},
	pages={381--386},
	year={2013},
	publisher={Taylor \& Francis}
}

@article{CowlsSchroeder2015,
	title={Causation, correlation, and big data in social science research},
	author={Cowls, Josh and Schroeder, Ralph},
	journal={Policy \& Internet},
	volume={7},
	number={4},
	pages={447--472},
	year={2015},
	publisher={Wiley Online Library}
}

@article{Guoetal2021,
	title={The effect of information disclosure on industry payments to physicians},
	author={Guo, Tong and Sriram, S and Manchanda, Puneet},
	journal={Journal of Marketing Research},
	volume={58},
	number={1},
	pages={115--140},
	year={2021},
	publisher={SAGE Publications Sage CA: Los Angeles, CA}
}

@article{GolderMacy2014,
	title={Digital footprints: Opportunities and challenges for online social research},
	author={Golder, Scott A and Macy, Michael W},
	journal={Annual Review of Sociology},
	volume={40},
	pages={129--152},
	year={2014},
	publisher={Annual Reviews}
}

@article{ZhangLuo2021,
	title={Can Consumer-Posted Photos Serve as a Leading Indicator of Restaurant Survival? Evidence from Yelp},
	author={Zhang, Mengxia and Luo, Lan},
	year={2021}
}

@techreport{Narangetal2019,
	title={The Impact of Mobile App Failures on Purchases in Online and Offline Channels},
	author={Narang, Unnati and Shankar, Venkatesh and Narayanan, Sridhar},
	year={2019},
	institution={Working Paper}
}

@article{Gordonetal2022,
	title={Close Enough? A Large-Scale Exploration of Non-Experimental Approaches to Advertising Measurement},
	author={Gordon, Brett R and Moakler, Robert and Zettelmeyer, Florian},
	journal={arXiv preprint arXiv:2201.07055},
	year={2022}
}

@article{Hunermundetal2021,
	title={Causal Machine Learning and Business Decision Making},
	author={H{\"u}nermund, Paul and Kaminski, Jermain and Schmitt, Carla},
	journal={Available at SSRN 3867326},
	year={2021}
}

@article{Quetal2021,
	title={Efficient Treatment Effect Estimation in Observational Studies under Heterogeneous Partial Interference},
	author={Qu, Zhaonan and Xiong, Ruoxuan and Liu, Jizhou and Imbens, Guido},
	journal={arXiv preprint arXiv:2107.12420},
	year={2021}
}

@article{TchetgenVanderWeele2012,
	title={On causal inference in the presence of interference},
	author={Tchetgen, Eric J Tchetgen and VanderWeele, Tyler J},
	journal={Statistical methods in medical research},
	volume={21},
	number={1},
	pages={55--75},
	year={2012},
	publisher={Sage Publications Sage UK: London, England}
}

@article{AtheyWager2021,
	title={Policy learning with observational data},
	author={Athey, Susan and Wager, Stefan},
	journal={Econometrica},
	volume={89},
	number={1},
	pages={133--161},
	year={2021},
	publisher={Wiley Online Library}
}

@Manual{R2022,
title = {R: A Language and Environment for Statistical Computing },
author = {{R Core Team}},
organization = {R Foundation for Statistical Computing},
 address = {Vienna, Austria},
year = {2022},
url = {https://www.R-project.org/},
  }

@article{Sverdrupetal2020,
	title={policytree: Policy learning via doubly robust empirical welfare maximization over trees},
	author={Sverdrup, Erik and Kanodia, Ayush and Zhou, Zhengyuan and Athey, Susan and Wager, Stefan},
	journal={Journal of Open Source Software},
	volume={5},
	number={50},
	pages={2232},
	year={2020}
}

@article{HuberSteinmayr2021,
author = {Martin Huber and Andreas Steinmayr},
title = {A Framework for Separating Individual-Level Treatment Effects From Spillover Effects},
journal = {Journal of Business \& Economic Statistics},
volume = {39},
pages = {422-436},
year  = {2021},
}

@article{AronowSamii2017,
author = {Peter M. Aronow and Cyrus Samii},
title = {Estimating average causal effects under general interference, with application to a social network experiment},
volume = {11},
journal = {The Annals of Applied Statistics},
pages = {1912-1947},
year = {2017},
}

@ARTICLE{KitagawaTetenov2018,
  author =       {T Kitagawa  and  A  Tetenov},
  title =        {Who  should  be  treated?   Empirical  welfare  maximization methods for treatment choice},
  journal =      {Econometrica},
  year =         {2018},
  volume =       {86},
  pages =        {591-616},
}

@ARTICLE{Stoye2009,
  author =       {J. Stoye},
  title =        {Minimax regret treatment choice with finite samples},
  journal =      {Journal  of  Econometrics},
  year =         {2009},
  volume =       {151},
  pages =        {70-81},
}

@ARTICLE{HiranoPorter2008,
  author =       {Hirano, K. and Porter, J.R.},
  title =        {Asymptotics for statistical treatment rules},
  journal =      {Econometrica},
  year=          {2009},
  volume = {77},
  pages = {1683-1701},
}

@ARTICLE{Manski2004,
  author =       {C F Manski},
  title =        {Statistical Treatment Rules for Heterogeneous Populations},
  journal =      {Econometrica},
  year =         {2004},
  volume =       {72},
  pages =        {1221-1246},
}

@article{SemenovaChernozhukov2021,
	title={Debiased machine learning of conditional average treatment effects and other causal functions},
	author={Semenova, Vira and Chernozhukov, Victor},
	journal={The Econometrics Journal},
	volume={24},
	number={2},
	pages={264--289},
	year={2021},
	publisher={Oxford University Press}
}

@article{kaggle2021,
	title = {Predicting Coupon Redemption},
	author = {Kaggle},
	year = {2019},
	journal = {https://www.kaggle.com/vasudeva009/predicting-coupon-redemption}
}

@techreport{Xingetal2020,
	title={"Quick Response" Economic Stimulus: The Effect of Small-Value Digital Coupons on Spending},
	author={Xing, Jianwei and Zou, Eric and Yin, Zhentao and Wang, Yong and Li, Zhenhua},
	year={2020},
	institution={National Bureau of Economic Research}
}

@article{Hsieh2010,
	title={Did Japan's shopping coupon program increase spending?},
	author={Hsieh, Chang-Tai and Shimizutani, Satoshi and Hori, Masahiro},
	journal={Journal of Public Economics},
	volume={94},
	number={7-8},
	pages={523--529},
	year={2010},
	publisher={Elsevier}
}

@article{grfmanual2022,
	title = {generalized random forests (grf 2.1.0)},
	author = {Athey, Susan and Friedberg, Rina and Hadad, Vitor and Hirshberg, David and Miner, Luke and Sverdrup, Erik and Tibshirani, Julie and Wager,Stefan and Wright, Marvin},
	year = {2022},
	journal = {https://grf-labs.github.io/grf/index.html},

}

@article{Robinson1988,
	title={Root-N-consistent semiparametric regression},
	author={Robinson, Peter M},
	journal={Econometrica: Journal of the Econometric Society},
	pages={931--954},
	year={1988},
	publisher={JSTOR}
}

@article{Smithetal2021,
	title={Optimal Price Targeting},
	author={Smith, Adam N and Seiler, Stephan and Aggarwal, Ishant},
	journal={Available at SSRN 3975957},
	year={2021}
}

@article{Holland86,
	title={Statistics and causal inference},
	author={Holland, Paul W},
	journal={Journal of the American statistical Association},
	volume={81},
	number={396},
	pages={945--960},
	year={1986},
	publisher={Taylor \& Francis}
}

@article{Cagalaetal2021,
	title={Optimal Targeting in Fundraising: A Causal Machine-Learning Approach},
	author={Cagala, Tobias and Glogowsky, Ulrich and Rincke, Johannes and Strittmatter, Anthony},
	journal={arXiv preprint arXiv:2103.10251},
	year={2021}
}

@article{AtheyWager2019,
	title={Estimating treatment effects with causal forests: An application},
	author={Athey, Susan and Wager, Stefan},
	journal={Observational Studies},
	volume={5},
	number={2},
	pages={37--51},
	year={2019},
	publisher={University of Pennsylvania Press}
}

@article{GlynnQuinn2010,
	title={An introduction to the augmented inverse propensity weighted estimator},
	author={Glynn, Adam N and Quinn, Kevin M},
	journal={Political analysis},
	volume={18},
	number={1},
	pages={36--56},
	year={2010},
	publisher={Cambridge University Press}
}

\pagebreak

\renewcommand\appendix{\par
	\setcounter{section}{0}%
	\setcounter{table}{0}%
	\setcounter{figure}{0}%
	\renewcommand\thesection{\Alph{section}}%
	\renewcommand\thetable{\Alph{section}.\arabic{table}}}
\renewcommand\thefigure{\Alph{section}.\arabic{figure}}
\clearpage

\numberwithin{equation}{section}
\noindent \textbf{\LARGE Appendices}

\begin{appendix}

	\section{Descriptive Statistics}\label{app:desc}
	\begin{table}[htp]
		\captionsetup{font=footnotesize}
		\begin{singlespace}
			\centering
			\adjustbox{width = 1.05\textwidth, center}{%
				\begin{tabular}{lcccccc}
					\toprule
					& \multicolumn{2}{c}{ready-to-eat food coupons} & \multicolumn{2}{c}{meat/seafood coupons} & \multicolumn{2}{c}{other food coupons} \\
					variable & received & not received & received & not received & received & not received \\
					\midrule
					daily expenditures & 237 & 257 & 234 & 264.37 & 241 & 335 \\ 
					age: 18-25 & 0.03 & 0.032 & 0.029 & 0.034 & 0.031 & 0.033 \\ 
					\ \ \ \ \ \ \ 26-35 & 0.101 & 0.103 & 0.092 & 0.12 & 0.099 & 0.159 \\ 
					\ \ \ \ \ \ \ 36-45 & 0.14 & 0.143 & 0.138 & 0.147 & 0.142 & 0.132 \\ 
					\ \ \ \ \ \ \ 46-55 & 0.191 & 0.192 & 0.191 & 0.192 & 0.191 & 0.208 \\ 
					\ \ \ \ \ \ \ 56-70 & 0.051 & 0.036 & 0.044 & 0.047 & 0.046 & 0.029 \\ 
					\ \ \ \ \ \ \ 70+ & 0.042 & 0.034 & 0.042 & 0.032 & 0.04 & 0.021 \\ 
					\ \ \ \ \ \ \ unknown & 0.444 & 0.461 & 0.463 & 0.427 & 0.452 & 0.418 \\ 
					family size: 1 & 0.174 & 0.167 & 0.173 & 0.169 & 0.172 & 0.164 \\ 
					\ \ \ \ \ \ \ \ \ \ \ \ \ \ \ \ 2 & 0.222 & 0.199 & 0.214 & 0.212 & 0.215 & 0.175 \\ 
					\ \ \ \ \ \ \ \ \ \ \ \ \ \ \ \ 3 & 0.079 & 0.078 & 0.077 & 0.083 & 0.078 & 0.092 \\ 
					\ \ \ \ \ \ \ \ \ \ \ \ \ \ \ \ 4 & 0.036 & 0.045 & 0.034 & 0.05 & 0.038 & 0.073 \\ 
					\ \ \ \ \ \ \ \ \ \ \ \ \ \ \ \ 5+ & 0.044 & 0.05 & 0.039 & 0.06 & 0.045 & 0.078 \\ 
					\ \ \ \ \ \ \ \ \ \ \ \ \ \ \ \ unknown & 0.444 & 0.461 & 0.463 & 0.427 & 0.452 & 0.418 \\ 
					marital status: married & 0.239 & 0.227 & 0.227 & 0.246 & 0.235 & 0.207 \\ 
					\ \ \ \ \ \ \ \ \ \ \ \ \ \ \ \ \ \ \ \  unmarried & 0.082 & 0.088 & 0.081 & 0.09 & 0.084 & 0.097 \\ 
					\ \ \ \ \ \ \ \ \ \ \ \ \ \ \ \ \ \ \ \  unknown & 0.679 & 0.686 & 0.691 & 0.664 & 0.681 & 0.696 \\ 
					dwelling type: rented & 0.031 & 0.036 & 0.029 & 0.041 & 0.032 & 0.06 \\ 
					\ \ \ \ \ \ \ \ \ \ \ \ \ \ \ \ \ \ \  \ owned & 0.525 & 0.503 & 0.507 & 0.532 & 0.516 & 0.521 \\ 
					\ \ \ \ \ \ \ \ \ \ \ \ \ \ \ \ \ \ \  \ unknown & 0.444 & 0.461 & 0.463 & 0.427 & 0.452 & 0.418 \\ 
					income group: 1 & 0.045 & 0.038 & 0.041 & 0.044 & 0.043 & 0.022 \\ 
					\ \ \ \ \ \ \ \ \ \ \ \ \ \ \ \ \ \ \ \  2 & 0.049 & 0.054 & 0.049 & 0.053 & 0.05 & 0.057 \\ 
					\ \ \ \ \ \ \ \ \ \ \ \ \ \ \ \ \ \ \ \  3 & 0.052 & 0.044 & 0.048 & 0.05 & 0.049 & 0.041 \\ 
					\ \ \ \ \ \ \ \ \ \ \ \ \ \ \ \ \ \ \ \  4 & 0.115 & 0.11 & 0.109 & 0.121 & 0.111 & 0.164 \\ 
					\ \ \ \ \ \ \ \ \ \ \ \ \ \ \ \ \ \ \ \  5 & 0.14 & 0.133 & 0.138 & 0.135 & 0.137 & 0.14 \\ 
					\ \ \ \ \ \ \ \ \ \ \ \ \ \ \ \ \ \ \ \  6 & 0.059 & 0.066 & 0.061 & 0.063 & 0.062 & 0.052 \\ 
					\ \ \ \ \ \ \ \ \ \ \ \ \ \ \ \ \ \ \ \  7 & 0.023 & 0.023 & 0.025 & 0.02 & 0.023 & 0.016 \\ 
					\ \ \ \ \ \ \ \ \ \ \ \ \ \ \ \ \ \ \ \  8 & 0.028 & 0.033 & 0.025 & 0.038 & 0.03 & 0.038 \\ 
					\ \ \ \ \ \ \ \ \ \ \ \ \ \ \ \ \ \ \ \  9 & 0.024 & 0.024 & 0.021 & 0.029 & 0.023 & 0.043 \\ 
					\ \ \ \ \ \ \ \ \ \ \ \ \ \ \ \ \ \ \ \  10 & 0.008 & 0.003 & 0.006 & 0.008 & 0.006 & 0.008 \\ 
					\ \ \ \ \ \ \ \ \ \ \ \ \ \ \ \ \ \ \ \  11 & 0.003 & 0.003 & 0.003 & 0.002 & 0.003 & 0 \\ 
					\ \ \ \ \ \ \ \ \ \ \ \ \ \ \ \ \ \ \ \  12 & 0.011 & 0.008 & 0.01 & 0.01 & 0.01 & 0 \\ 
					\ \ \ \ \ \ \ \ \ \ \ \ \ \ \ \ \ \ \ \  unknown & 0.444 & 0.461 & 0.463 & 0.427 & 0.452 & 0.418 \\ 
					coupons redeemed & 0.037 & 0.018 & 0.036 & 0.018 & 0.031 & 0.006 \\
					\bottomrule
			\end{tabular}}
			\caption[Descriptive statistics, variable means among food coupon (non-)recipients]{Mean of the variables among the treated who received a coupon of a certain category and of those who did not}
		\end{singlespace}
	\end{table}
	
	\begin{table}[htp]
		\captionsetup{font=footnotesize}
		\begin{singlespace}
			\centering
			\adjustbox{width = 0.8\textwidth, center}{%
				\begin{tabular}{lcccc}
					\toprule
					& \multicolumn{2}{c}{drugstore coupons} & \multicolumn{2}{c}{other  non-food coupons} \\
					variable & received & not received & received & not received \\
					\midrule
					daily expenditures & 243 & 272 & 246 & 243.54 \\ 
					age: 18-25 & 0.031 & 0.033 & 0.036 & 0.028 \\ 
					\ \ \ \ \ \ \ 26-35 & 0.104 & 0.07 & 0.107 & 0.099 \\ 
					\ \ \ \ \ \ \ 36-45 & 0.143 & 0.117 & 0.149 & 0.137 \\ 
					\ \ \ \ \ \ \ 46-55 & 0.191 & 0.198 & 0.2 & 0.187 \\ 
					\ \ \ \ \ \ \ 56-70 & 0.046 & 0.035 & 0.047 & 0.044 \\ 
					\ \ \ \ \ \ \ 70+ & 0.039 & 0.033 & 0.04 & 0.038 \\ 
					\ \ \ \ \ \ \ unknown & 0.447 & 0.514 & 0.422 & 0.466 \\ 
					family size: 1 & 0.172 & 0.162 & 0.177 & 0.168 \\ 
					\ \ \ \ \ \ \ \ \ \ \ \ \ \ \ \ 2 & 0.215 & 0.176 & 0.224 & 0.207 \\ 
					\ \ \ \ \ \ \ \ \ \ \ \ \ \ \ \ 3 & 0.079 & 0.076 & 0.079 & 0.079 \\ 
					\ \ \ \ \ \ \ \ \ \ \ \ \ \ \ \ 4 & 0.04 & 0.031 & 0.043 & 0.038 \\ 
					\ \ \ \ \ \ \ \ \ \ \ \ \ \ \ \ 5+ & 0.047 & 0.04 & 0.055 & 0.042 \\ 
					\ \ \ \ \ \ \ \ \ \ \ \ \ \ \ \ unknown & 0.447 & 0.514 & 0.422 & 0.466 \\ 
					marital status: married & 0.236 & 0.205 & 0.257 & 0.221 \\ 
					\ \ \ \ \ \ \ \ \ \ \ \ \ \ \ \ \ \ \ \  unmarried & 0.085 & 0.074 & 0.075 & 0.09 \\ 
					\ \ \ \ \ \ \ \ \ \ \ \ \ \ \ \ \ \ \ \  unknown & 0.68 & 0.721 & 0.668 & 0.689 \\ 
					dwelling type: rented & 0.033 & 0.047 & 0.032 & 0.034 \\ 
					\ \ \ \ \ \ \ \ \ \ \ \ \ \ \ \ \ \ \  \ owned & 0.521 & 0.438 & 0.547 & 0.499 \\ 
					\ \ \ \ \ \ \ \ \ \ \ \ \ \ \ \ \ \ \  \ unknown & 0.447 & 0.514 & 0.422 & 0.466 \\ 
					income group: 1 & 0.041 & 0.05 & 0.055 & 0.035 \\ 
					\ \ \ \ \ \ \ \ \ \ \ \ \ \ \ \ \ \ \ \  2 & 0.051 & 0.047 & 0.05 & 0.051 \\ 
					\ \ \ \ \ \ \ \ \ \ \ \ \ \ \ \ \ \ \ \  3 & 0.05 & 0.037 & 0.046 & 0.051 \\ 
					\ \ \ \ \ \ \ \ \ \ \ \ \ \ \ \ \ \ \ \  4 & 0.114 & 0.1 & 0.115 & 0.112 \\ 
					\ \ \ \ \ \ \ \ \ \ \ \ \ \ \ \ \ \ \ \  5 & 0.139 & 0.114 & 0.149 & 0.131 \\ 
					\ \ \ \ \ \ \ \ \ \ \ \ \ \ \ \ \ \ \ \  6 & 0.062 & 0.057 & 0.065 & 0.059 \\ 
					\ \ \ \ \ \ \ \ \ \ \ \ \ \ \ \ \ \ \ \  7 & 0.023 & 0.018 & 0.022 & 0.023 \\ 
					\ \ \ \ \ \ \ \ \ \ \ \ \ \ \ \ \ \ \ \  8 & 0.031 & 0.017 & 0.031 & 0.029 \\ 
					\ \ \ \ \ \ \ \ \ \ \ \ \ \ \ \ \ \ \ \  9 & 0.024 & 0.027 & 0.022 & 0.025 \\ 
					\ \ \ \ \ \ \ \ \ \ \ \ \ \ \ \ \ \ \ \  10 & 0.006 & 0.006 & 0.008 & 0.005 \\ 
					\ \ \ \ \ \ \ \ \ \ \ \ \ \ \ \ \ \ \ \  11 & 0.003 & 0.005 & 0.003 & 0.003 \\ 
					\ \ \ \ \ \ \ \ \ \ \ \ \ \ \ \ \ \ \ \  12 & 0.01 & 0.008 & 0.01 & 0.009 \\ 
					\ \ \ \ \ \ \ \ \ \ \ \ \ \ \ \ \ \ \ \  unknown & 0.447 & 0.514 & 0.422 & 0.466 \\ 
					coupons redeemed & 0.031 & 0.005 & 0.043 & 0.022 \\ 
					\bottomrule
			\end{tabular}}
			\caption[Descriptive statistics, variable means among non-food coupon (non-)recipients]{Mean of the variables among the treated who received a coupon of a certain category and of those who did not}
		\end{singlespace}
	\end{table}

	\begin{table}[htp]
		\captionsetup{font=footnotesize}
		\begin{singlespace}
			\centering
			\adjustbox{width = 0.9\textwidth, center}{%
				\begin{tabular}{lccccc}
					\toprule
					variable & Overall & Coupon Receivers & Non-Receivers & Diff & p-val \\
					N & 50,624 & 15,327 & 35,297 & &\\
					\midrule
					By Brand Type: &&&&&\\
					Established Brands & 221 & 255 & 206 & 48 & 0 \\ 
					Local Brands & 59.17 & 74.01 & 52.73 & 21.28 & 0 \\ 
					By Product Type: &&&&&\\
					Alcohol & 0.657 & 0.975 & 0.519 & 0.46 & 0 \\ 
					Bakery & 78.79 & 85.1 & 76.05 & 9.05 & 0 \\ 
					Dairy, Juices \& Snacks & 4.72 & 6.08 & 4.13 & 1.96 & 0 \\ 
					Flowers \& Plants & 0.699 & 0.846 & 0.635 & 0.21 & 0.02 \\ 
					Fuel & 94.35 & 107.2 & 88.77 & 18.42 & 0 \\ 
					Garden & 1.93 & 2.61 & 1.63 & 0.98 & 0 \\ 
					Grocery & 114 & 136 & 105 & 31 & 0 \\ 
					Meat & 82.04 & 88.45 & 79.25 & 9.2 & 0 \\ 
					Miscellaneous & 4.7 & 5.47 & 4.37 & 1.1 & 0 \\ 
					Natural Products & 6 & 7.11 & 5.51 & 1.6 & 0 \\ 
					Packaged Meat & 87.86 & 95.4 & 84.59 & 10.81 & 0 \\ 
					Pharmaceutical & 31.8 & 38.88 & 28.72 & 10.16 & 0 \\ 
					Prepared Food & 2.48 & 3.05 & 2.24 & 0.81 & 0 \\ 
					Restaurant & 76.08 & 81.94 & 73.54 & 8.41 & 0 \\ 
					Salads & 1.71 & 2.16 & 1.52 & 0.64 & 0 \\ 
					Seafood & 1.86 & 2.08 & 1.76 & 0.32 & 0.01 \\ 
					Skin \& Hair Care & 76.94 & 82.95 & 74.32 & 8.63 & 0 \\ 
					Travel & 1.7 & 2.14 & 1.5 & 0.64 & 0 \\ 
					Vegetables (cut) & 0.017 & 0.026 & 0.014 & 0.01 & 0.03 \\ 
					\bottomrule
			\end{tabular}}\caption[Daily expenditures by product type]{Mean of daily expenditures by brand and product type in the total sample ('Overall`), among coupon receivers and non-receivers as well as the mean difference across treatment states and the p-value of a two-sample t-test.}\label{tab:compositionexpenditures}
		\end{singlespace}
	\end{table}

	\FloatBarrier
	
	\goodbreak

	\section{Propensity Score Distribution}\label{appendix:propensity}
	\begin{figure}[H]
		\centering
		\makebox[\textwidth][c]{
			\begin{tabular}{cc}
				\includegraphics[width=.45\textwidth]{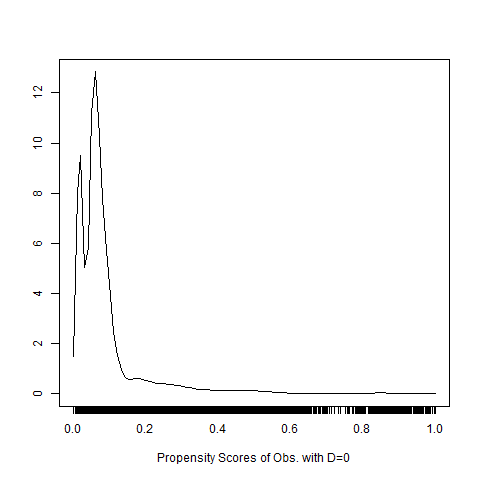}&
				\includegraphics[width=.45\textwidth]{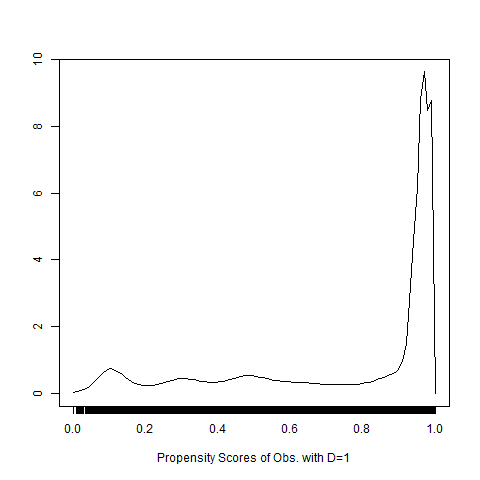}\\ \textbf{(a)}  & \textbf{(b)} \\
		\end{tabular}}
		\makebox[\textwidth][c]{
			\begin{tabular}{cc}
				\includegraphics[width=.45\textwidth]{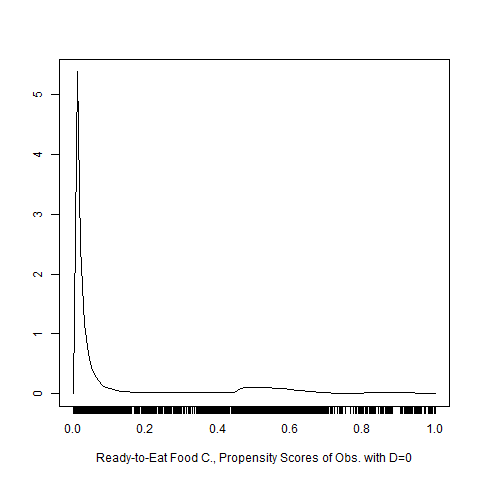}&
				\includegraphics[width=.45\textwidth]{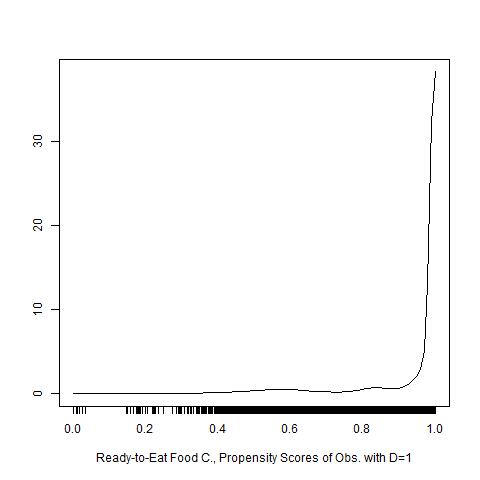} \\
				\textbf{(c)}  & \textbf{(d)} \\
		\end{tabular}}
		\caption[Propensity scores of receiving any coupon \& ready-to-eat food coupons]{Distribution of propensity scores of receiving any coupon among observations that received (a) any coupon and (b) no coupon, as well as that of the propensity scores of receiving ready-to-eat food coupons among observations that (c) did and (d) did not received ready-to-eat food coupons. The plots are produced with the \texttt{logspline} command in R with the lower and upper bounds of the support of the propensity scores are set to 0 and 1.}
		\label{figure:propAllReady}
	\end{figure}
	
	\begin{figure}
		\centering
		\makebox[\textwidth][c]{
			\begin{tabular}{cc}
				\includegraphics[width=.45\textwidth]{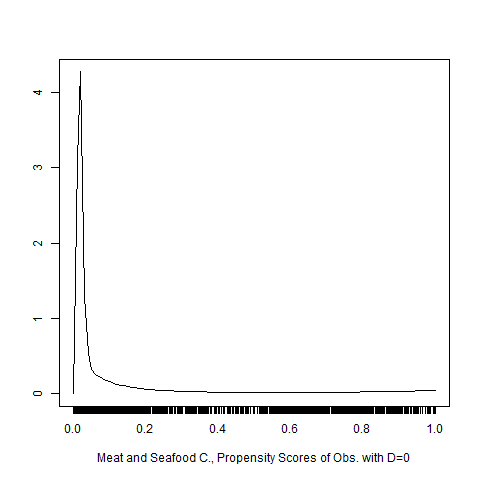}&
				\includegraphics[width=.45\textwidth]{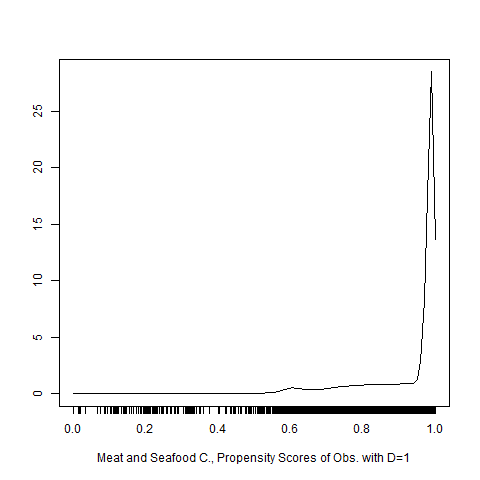} \\
				\textbf{(a)}  & \textbf{(b)} \\[5pt]
		\end{tabular}}
		\makebox[\textwidth][c]{
			\begin{tabular}{cc}
				\includegraphics[width=.45\textwidth]{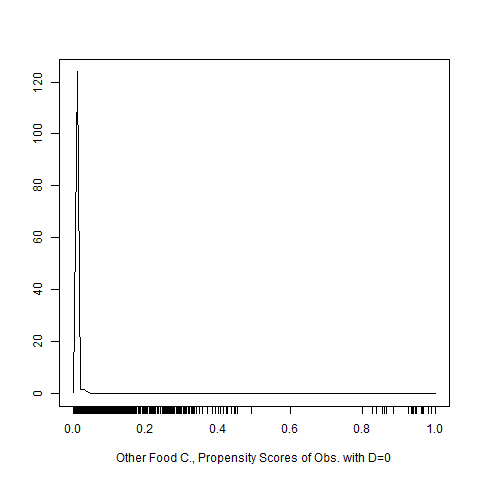}&
				\includegraphics[width=.45\textwidth]{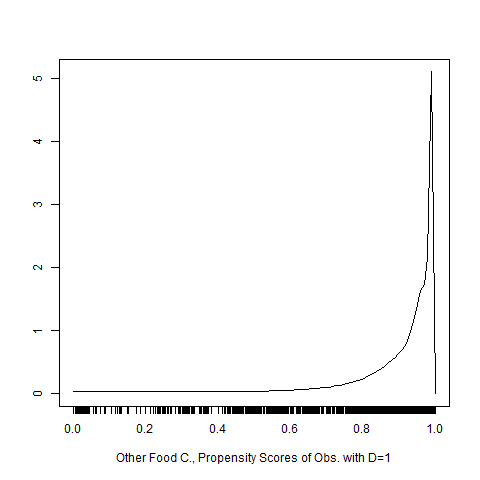} \\
				\textbf{(c)}  & \textbf{(d)} \\[5pt]
		\end{tabular}}
		\caption[Propensity scores of receiving meat/seafood coupon \& other food coupons]{Distribution of propensity scores of receiving meat/seafood coupons among observations that (c) did and (d) did not received meat/seafood coupons, as well as that of the propensity scores of receiving other food coupons among observations that (c) did and (d) did not received other food coupons. The plots are produced with the \texttt{logspline} command in R with the lower and upper bounds of the support of the propensity scores are set to 0 and 1.}
		\label{figure:propMeatOtherFood}
	\end{figure}
	
	\begin{figure}[htp]
		\centering
		\makebox[\textwidth][c]{
			\begin{tabular}{cc}
				\includegraphics[width=.45\textwidth]{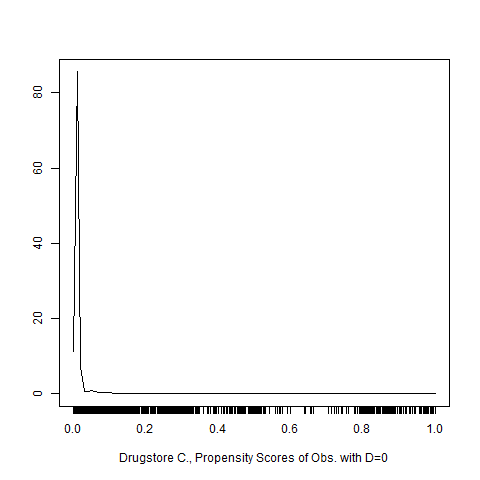}&
				\includegraphics[width=.45\textwidth]{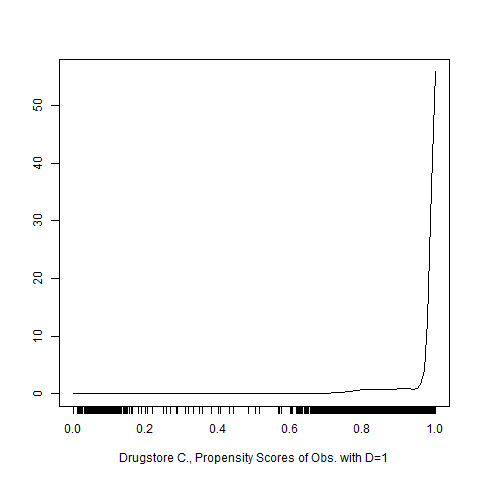} \\
				\textbf{(a)}  & \textbf{(b)} \\[5pt]
		\end{tabular}}
		\makebox[\textwidth][c]{
			\begin{tabular}{cc}
				\includegraphics[width=.45\textwidth]{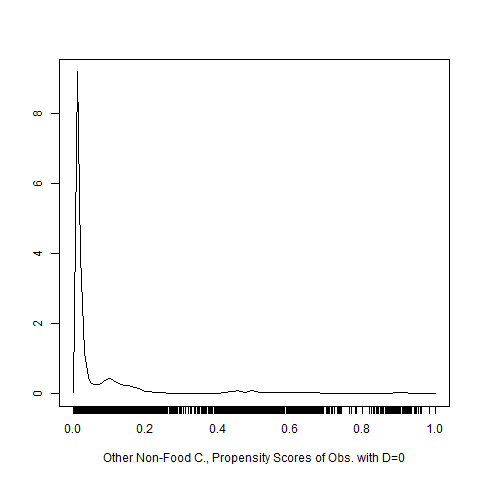}&
				\includegraphics[width=.45\textwidth]{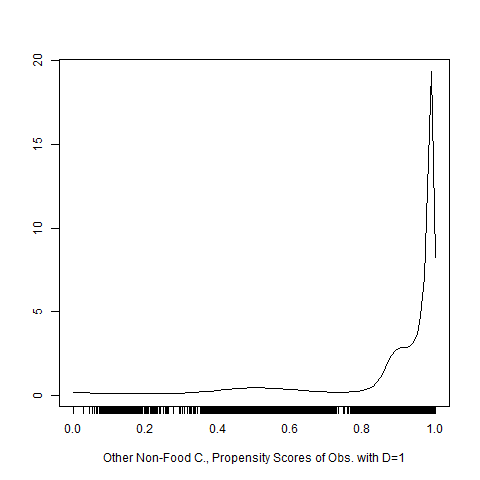} \\
				\textbf{(c)}  & \textbf{(d)} \\[5pt]
		\end{tabular}}
		\caption[Propensity scores of receiving drugstore coupons \& other non-food coupons]{Distribution of propensity scores of receiving drugstore coupons among observations that (c) did and (d) did not received drugstore coupons, as well as that of the propensity scores of receiving other non-food coupons among observations that (c) did and (d) did not received other non-food coupons. The plots are produced with the \texttt{logspline} command in R with the lower and upper bounds of the support of the propensity scores are set to 0 and 1.}
		\label{figure:propOther}
	\end{figure}
	
	\FloatBarrier
	\section{GATE Estimates for Coupons Applicable to Plants, Drugstore Items and Other Products}\label{appendix:GATE}
	
	\begin{figure}[H]
		\centering
		\makebox[\textwidth][c]{
			\begin{tabular}{cc}
				\includegraphics[width=.45\textwidth]{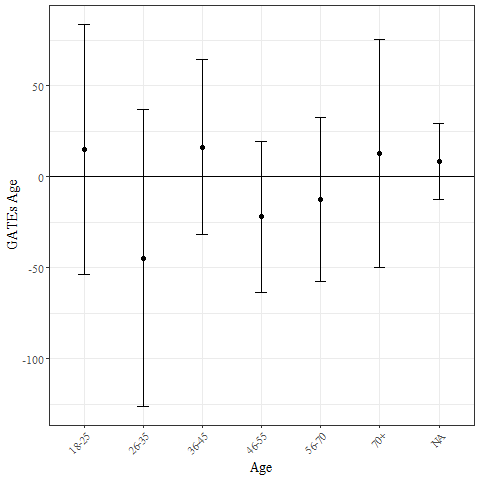}&
				\includegraphics[width=.45\textwidth]{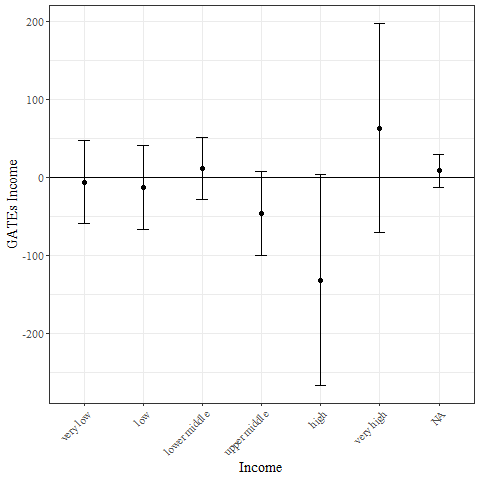} \\[25pt]
		\end{tabular}}
		\makebox[\textwidth][c]{
			\begin{tabular}{cc}
				\includegraphics[width=.45\textwidth]{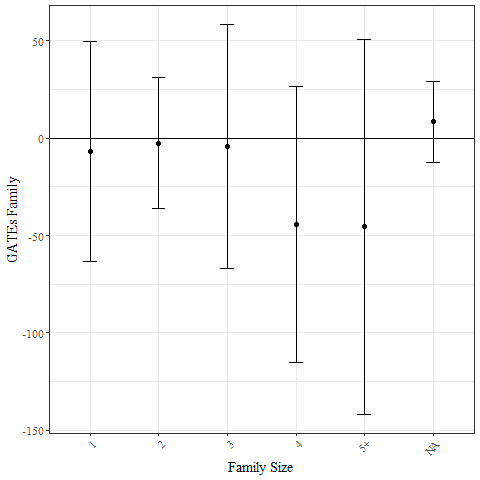}&
				\includegraphics[width=.45\textwidth]{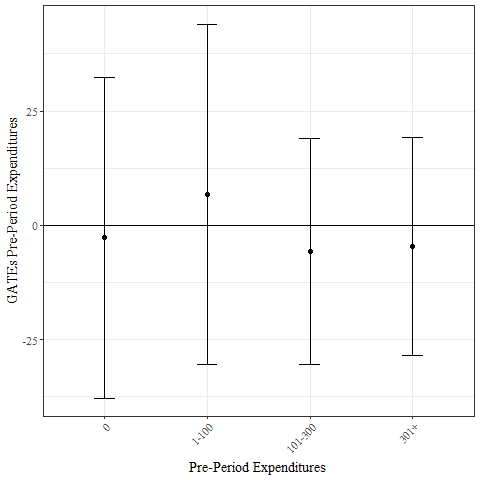} \\[25pt]
		\end{tabular}}
		\caption[GATEs of ready-to-eat food coupons]{GATEs of ready-to-eat food coupons with 95\% confidence interval.}
		\label{figure:GATEReadyEat}
	\end{figure}
	
	\begin{figure}[htp]
		\centering
		\makebox[\textwidth][c]{
			\begin{tabular}{cc}
				\includegraphics[width=.45\textwidth]{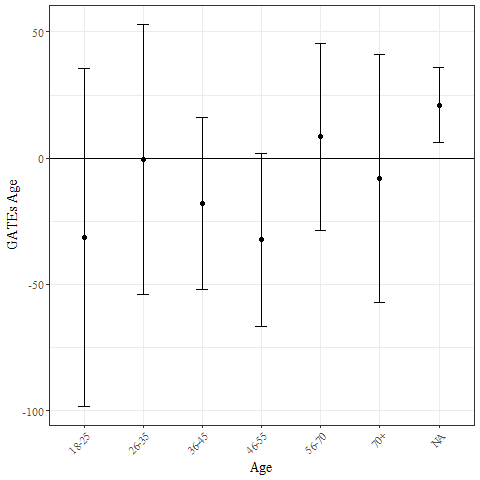}&
				\includegraphics[width=.45\textwidth]{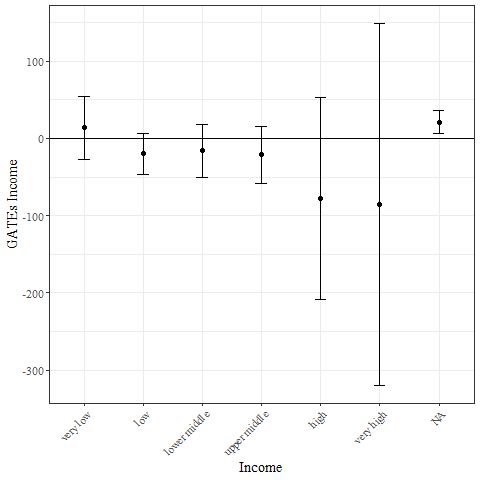} \\[25pt]
		\end{tabular}}
		\makebox[\textwidth][c]{
			\begin{tabular}{cc}
				\includegraphics[width=.45\textwidth]{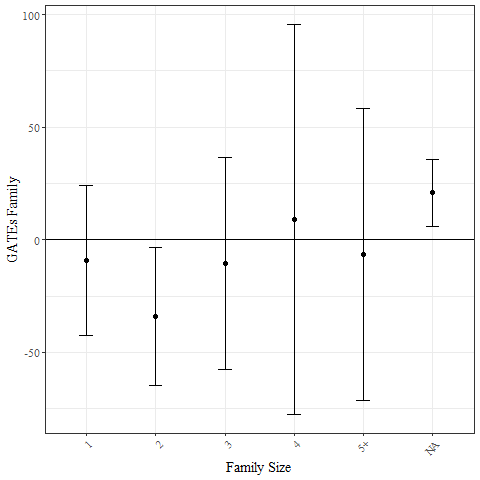}&
				\includegraphics[width=.45\textwidth]{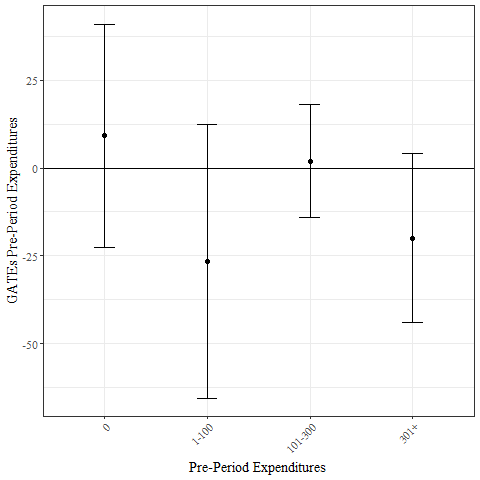} \\[25pt]
		\end{tabular}}
		\caption[GATEs of meat/seafood coupons]{GATEs of meat/seafood coupons with 95\% confidence interval.}
		\label{figure:GATEPlants}
	\end{figure}

	\begin{figure}[htp]
		\centering
		\makebox[\textwidth][c]{
			\begin{tabular}{cc}
				\includegraphics[width=.45\textwidth]{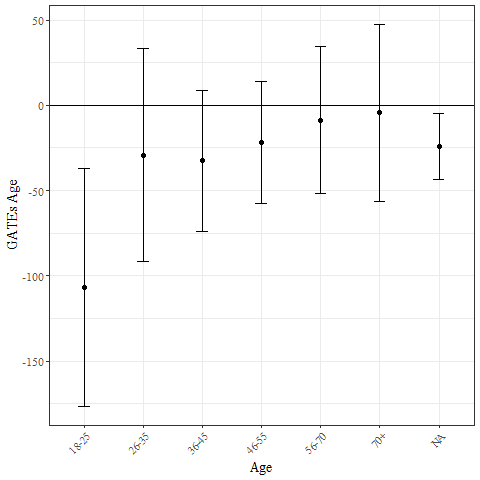}&
				\includegraphics[width=.45\textwidth]{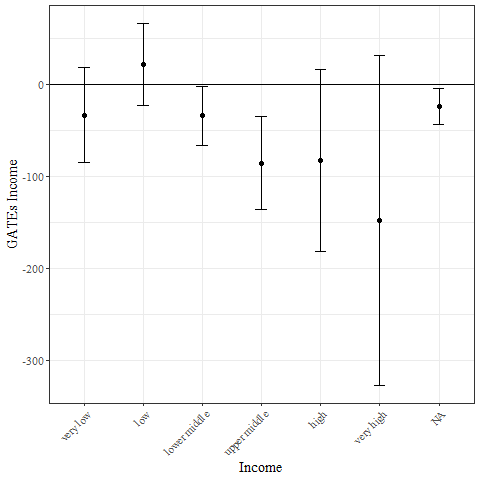} \\[25pt]
		\end{tabular}}
		\makebox[\textwidth][c]{
			\begin{tabular}{cc}
				\includegraphics[width=.45\textwidth]{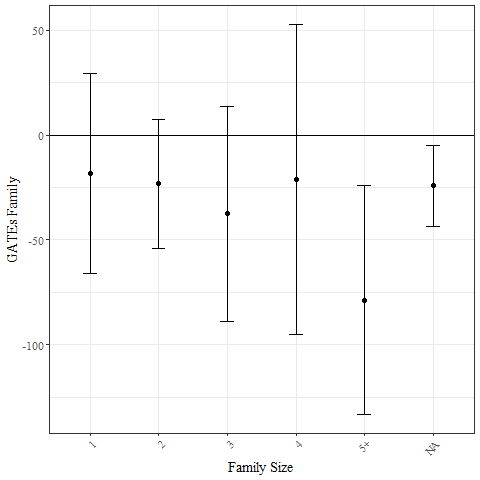}&
				\includegraphics[width=.45\textwidth]{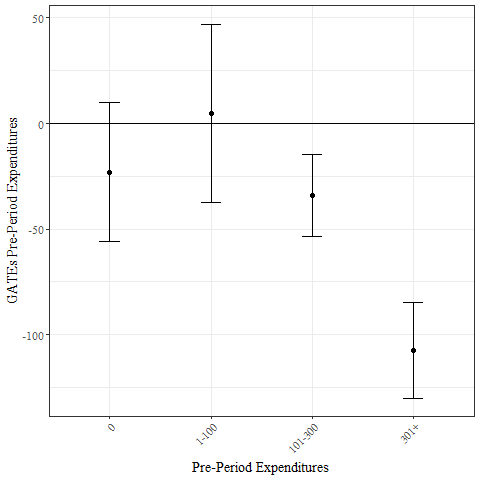} \\[25pt]
		\end{tabular}}
		\caption[GATEs of non-food coupons]{GATEs of coupons applicable to other non-food products with 95\% confidence interval.}
		\label{figure:GATEOther}
	\end{figure}
	
	\FloatBarrier
	\section{Robustness Checks}\label{appendix:Robustness}
	\subsection{Reduced Dataset: GATE Estimates}
	\begin{figure}[htp]
		\centering
		\makebox[\textwidth][c]{
			\begin{tabular}{cc}
				\includegraphics[width=.45\textwidth]{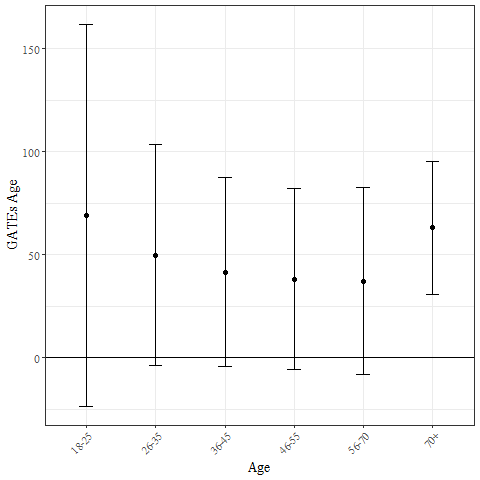}&
				\includegraphics[width=.45\textwidth]{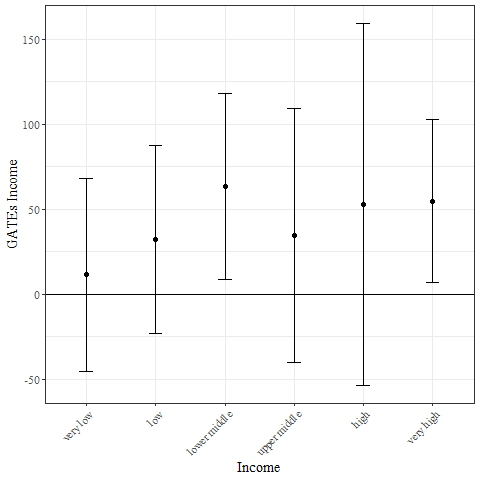} \\[25pt]
		\end{tabular}}
		\makebox[\textwidth][c]{
			\begin{tabular}{cc}
				\includegraphics[width=.45\textwidth]{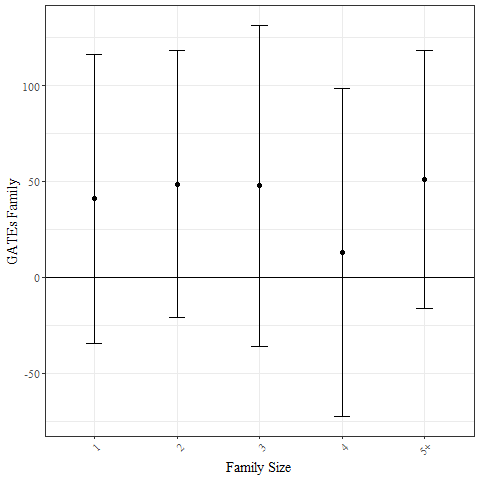}&
				\includegraphics[width=.45\textwidth]{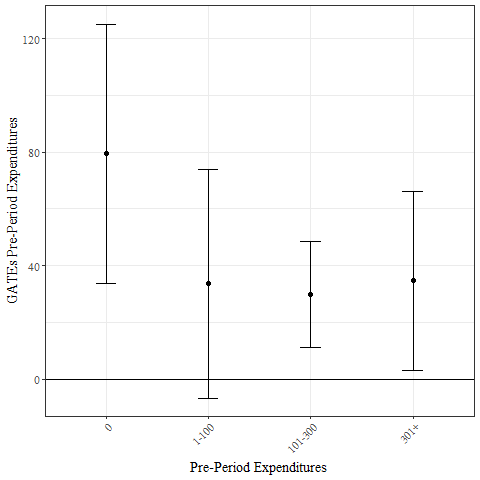} \\[25pt]
		\end{tabular}}
		\caption[GATEs of receiving any coupon, estimated in reduced dataset]{GATEs of receiving any coupon with 95\% confidence interval, estimated in reduced dataset.}
		\label{figure:GATE_robustness}
	\end{figure}
	
	\begin{figure}[htp]
		\centering
		\makebox[\textwidth][c]{
			\begin{tabular}{cc}
				\includegraphics[width=.45\textwidth]{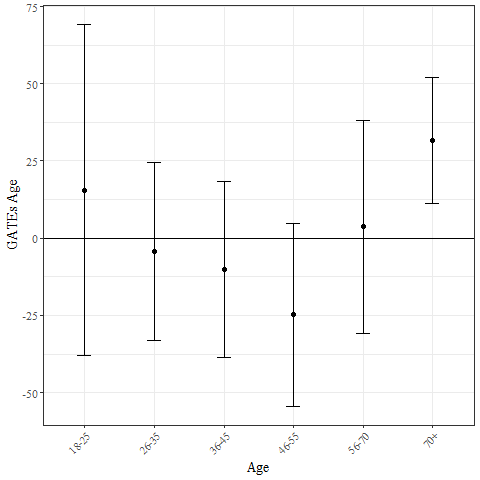}&
				\includegraphics[width=.45\textwidth]{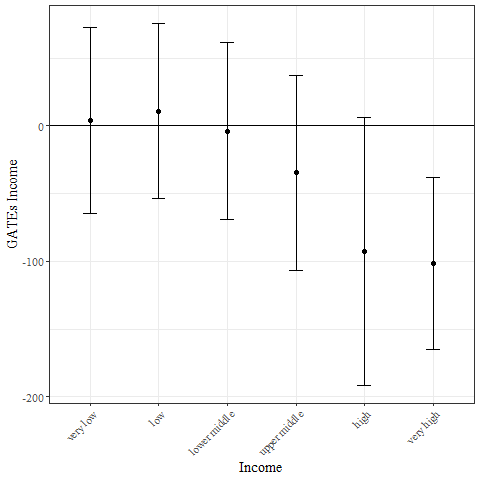} \\[25pt]
		\end{tabular}}
		\makebox[\textwidth][c]{
			\begin{tabular}{cc}
				\includegraphics[width=.45\textwidth]{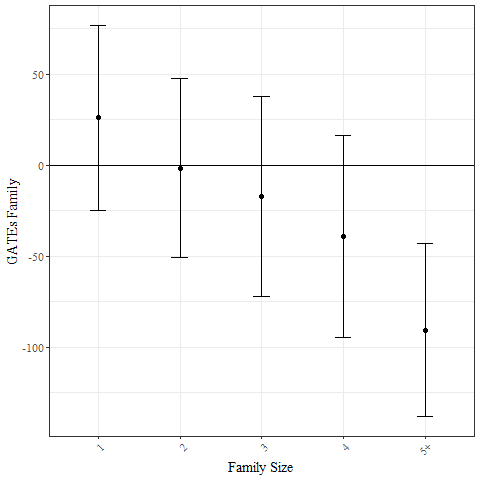}&
				\includegraphics[width=.45\textwidth]{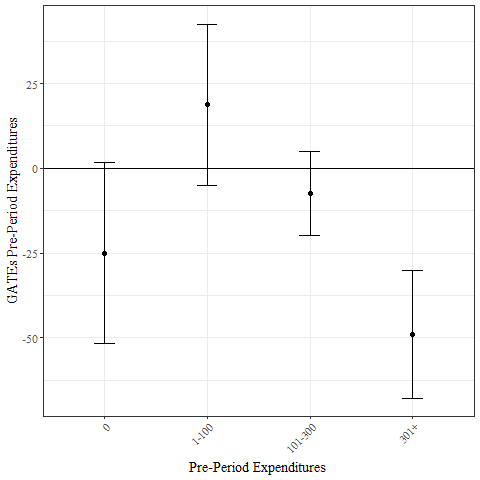} \\[25pt]
		\end{tabular}}
		\caption[GATEs of ready-to-eat food coupons, estimated in reduced dataset]{GATEs of ready-to-eat food coupons with 95\% confidence interval, estimated in reduced dataset.}
		\label{figure:GATEReady_robustness}
	\end{figure}
	
	\begin{figure}[htp]
		\centering
		\makebox[\textwidth][c]{
			\begin{tabular}{cc}
				\includegraphics[width=.45\textwidth]{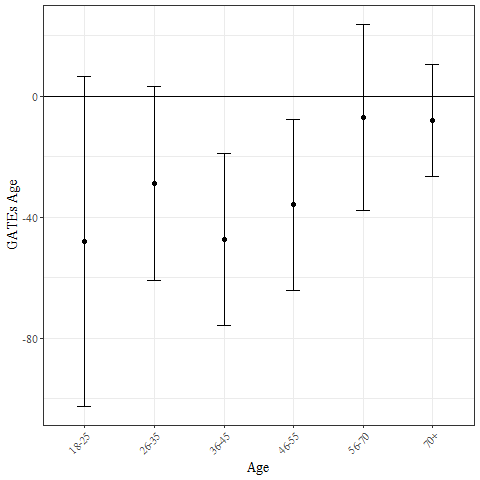}&
				\includegraphics[width=.45\textwidth]{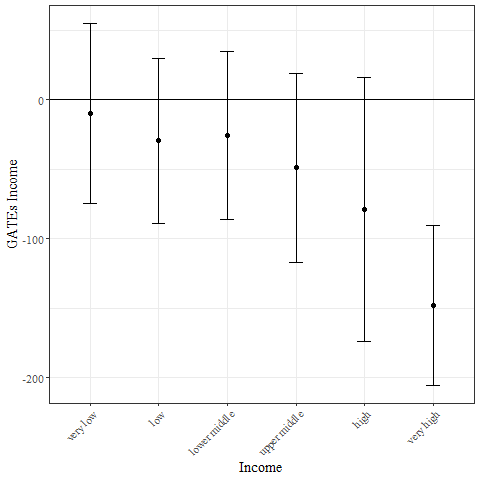} \\[25pt]
		\end{tabular}}
		\makebox[\textwidth][c]{
			\begin{tabular}{cc}
				\includegraphics[width=.45\textwidth]{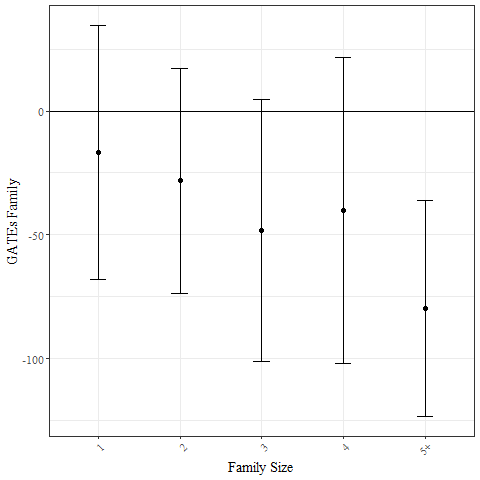}&
				\includegraphics[width=.45\textwidth]{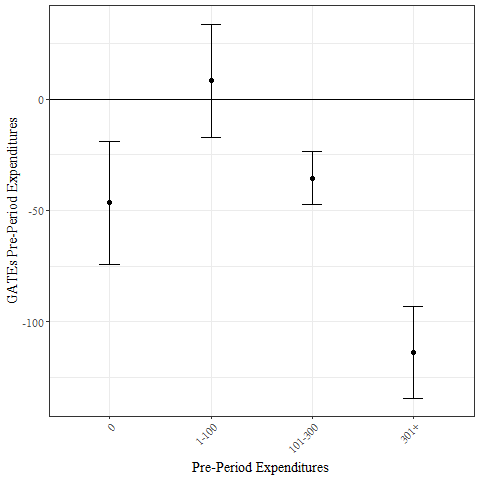} \\[25pt]
		\end{tabular}}
		\caption[GATEs of meat/seafood coupons, estimated in reduced dataset]{GATEs of meat and seafood coupons with 95\% confidence interval, estimated in reduced dataset.}
		\label{figure:GATEMeat_robustness}
	\end{figure}
	
	\begin{figure}[htp]
		\centering
		\makebox[\textwidth][c]{
			\begin{tabular}{cc}
				\includegraphics[width=.45\textwidth]{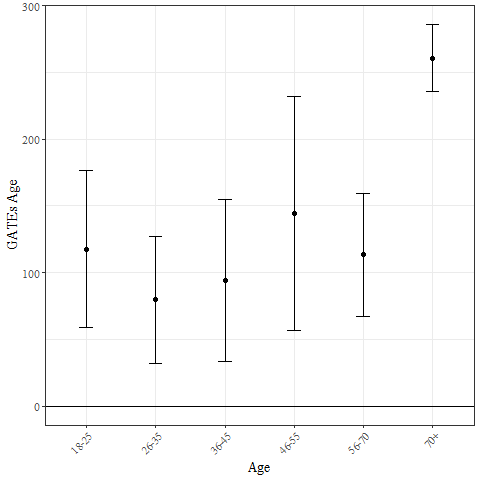}&
				\includegraphics[width=.45\textwidth]{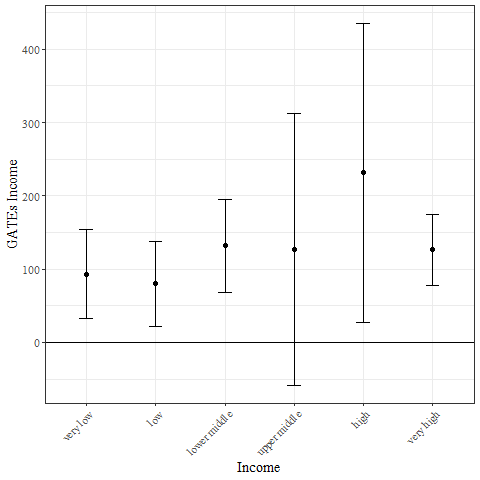} \\[25pt]
		\end{tabular}}
		\makebox[\textwidth][c]{
			\begin{tabular}{cc}
				\includegraphics[width=.45\textwidth]{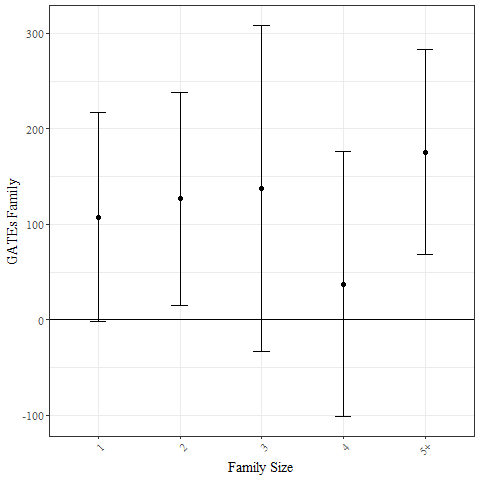}&
				\includegraphics[width=.45\textwidth]{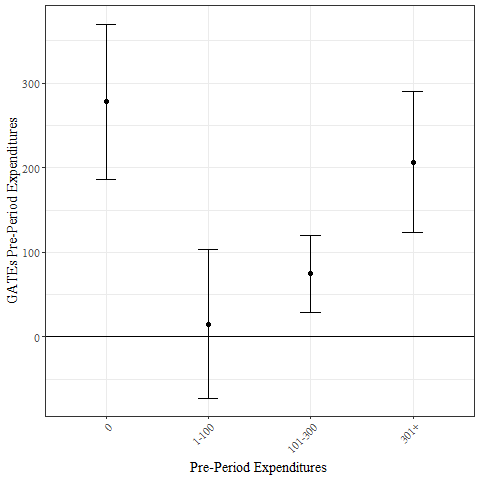} \\[25pt]
		\end{tabular}}
		\caption[GATEs of other food coupons, estimated in reduced dataset]{GATEs of coupons applicable to other food items with 95\% confidence interval, estimated in reduced dataset.}
		\label{figure:GATEOtherFood_robustness}
	\end{figure}
	
	\begin{figure}[htp]
		\centering
		\makebox[\textwidth][c]{
			\begin{tabular}{cc}
				\includegraphics[width=.45\textwidth]{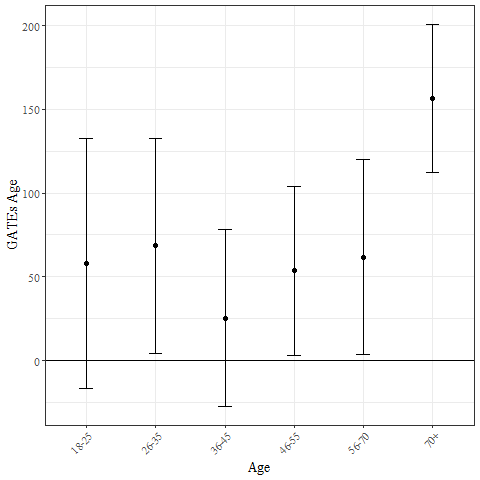}&
				\includegraphics[width=.45\textwidth]{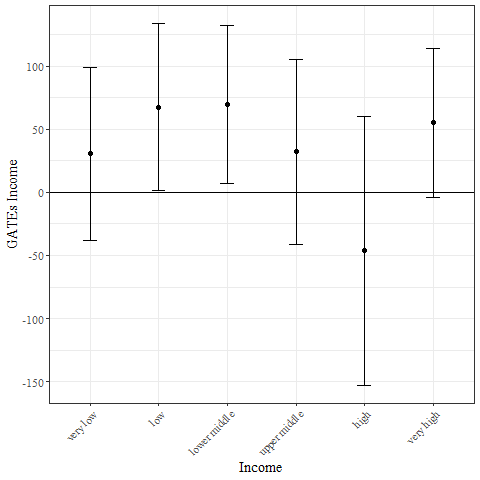} \\[25pt]
		\end{tabular}}
		\makebox[\textwidth][c]{
			\begin{tabular}{cc}
				\includegraphics[width=.45\textwidth]{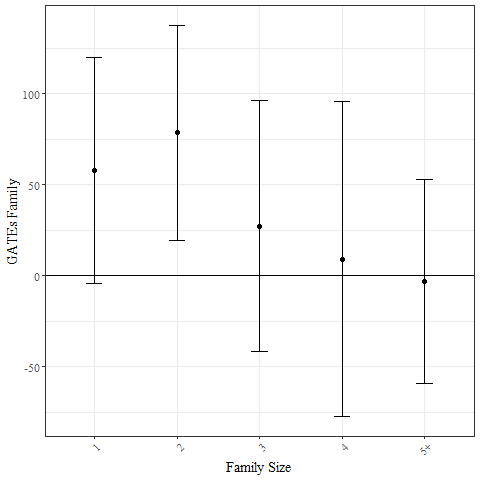}&
				\includegraphics[width=.45\textwidth]{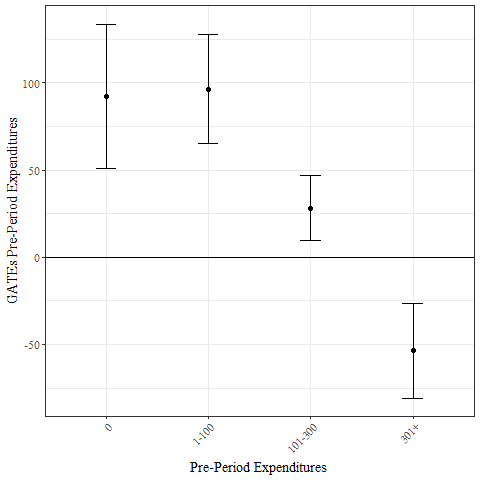} \\[25pt]
		\end{tabular}}
		\caption[GATEs of drugstore coupons, estimated in reduced dataset]{GATEs of drugstore coupons with 95\% confidence interval, estimated in reduced dataset.}
		\label{figure:GATEDrugstore_robustness}
	\end{figure}

	\begin{figure}[htp]
		\centering
		\makebox[\textwidth][c]{
			\begin{tabular}{cc}
				\includegraphics[width=.45\textwidth]{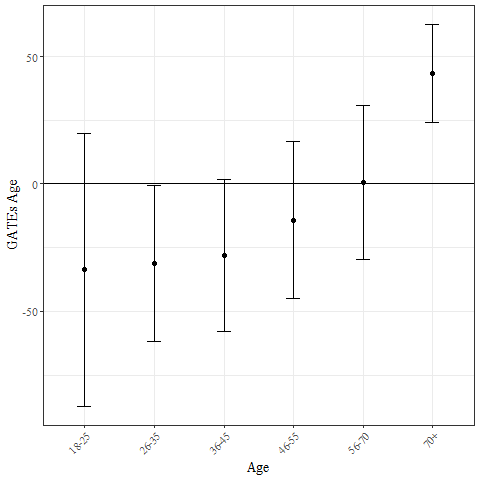}&
				\includegraphics[width=.45\textwidth]{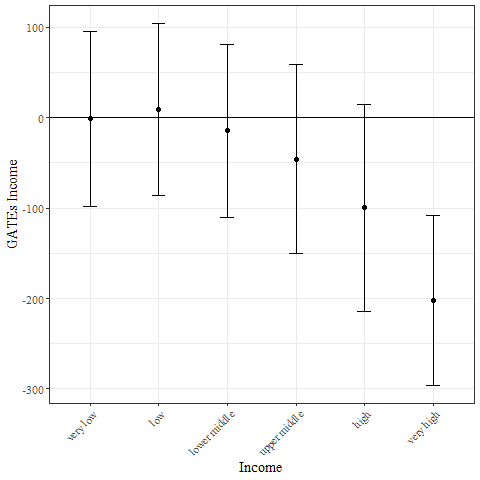} \\[25pt]
		\end{tabular}}
		\makebox[\textwidth][c]{
			\begin{tabular}{cc}
				\includegraphics[width=.45\textwidth]{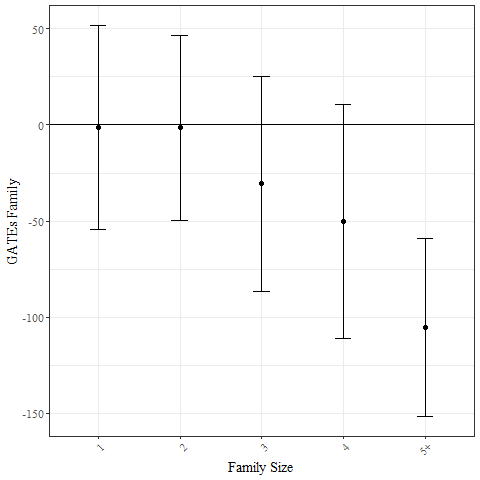}&
				\includegraphics[width=.45\textwidth]{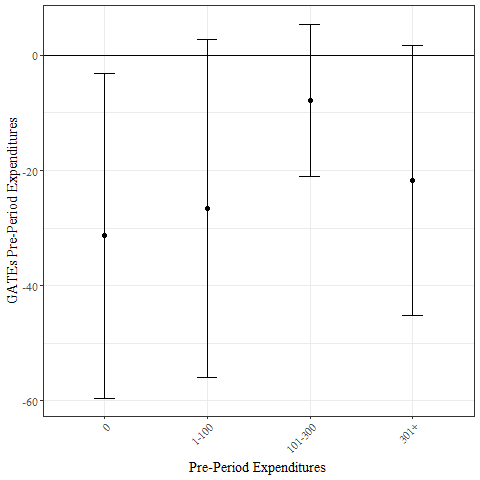} \\[25pt]
		\end{tabular}}
		\caption[GATEs of other non-food coupons, estimated in reduced dataset]{GATEs of coupons applicable to other non-food items with 95\% confidence interval, estimated in reduced dataset.}
		\label{figure:GATEOtherNonFood_robustness}
	\end{figure}
	
	\FloatBarrier
	\subsection{Reduced Dataset: Policy Trees}
	
	\begin{figure}[htp]
		\centering
		\makebox[\textwidth][c]{
			\begin{tabular}{cc}
				\includegraphics[width=.60\textwidth]{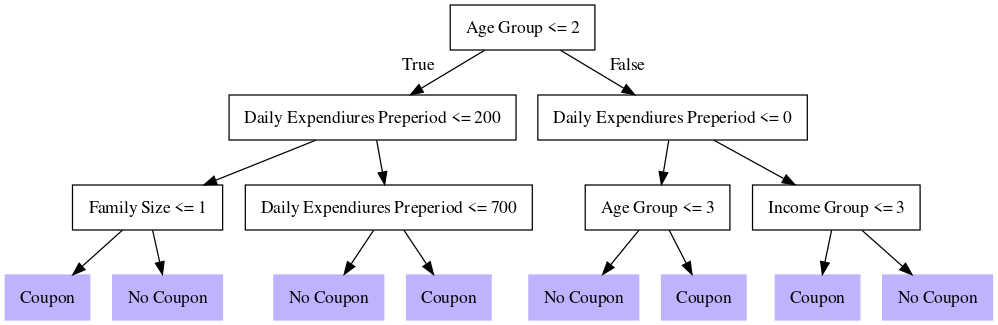}&
				\includegraphics[width=.6\textwidth]{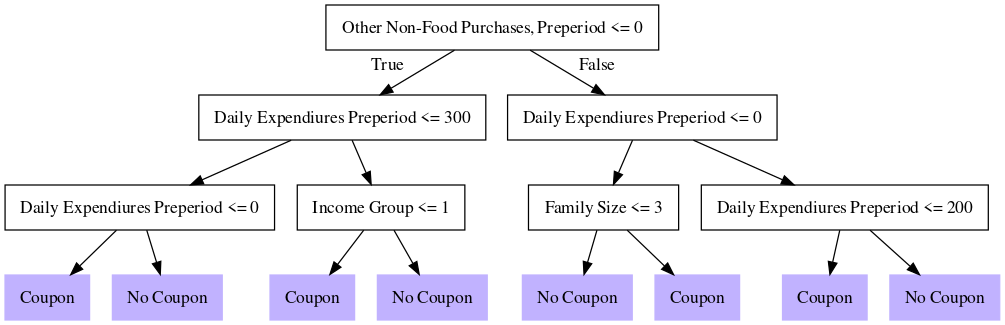} \\
				\textbf{(a)}  & \textbf{(b)} \\[25pt]
		\end{tabular}}
		\makebox[\textwidth][c]{
			\begin{tabular}{cc}
				\includegraphics[width=.6\textwidth]{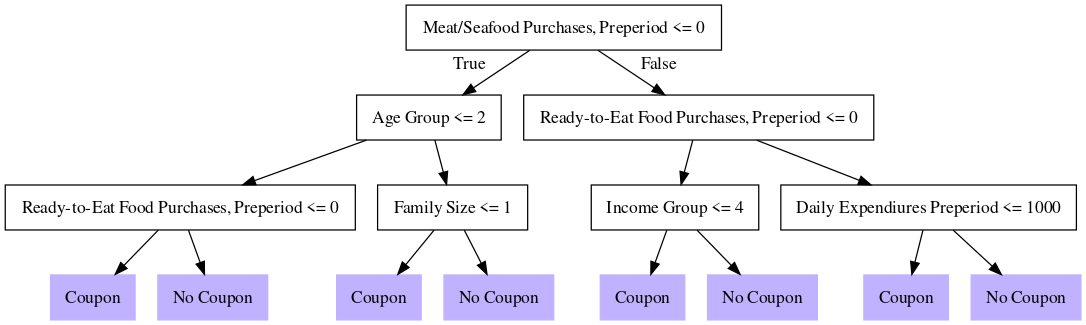}&
				\includegraphics[width=.6\textwidth]{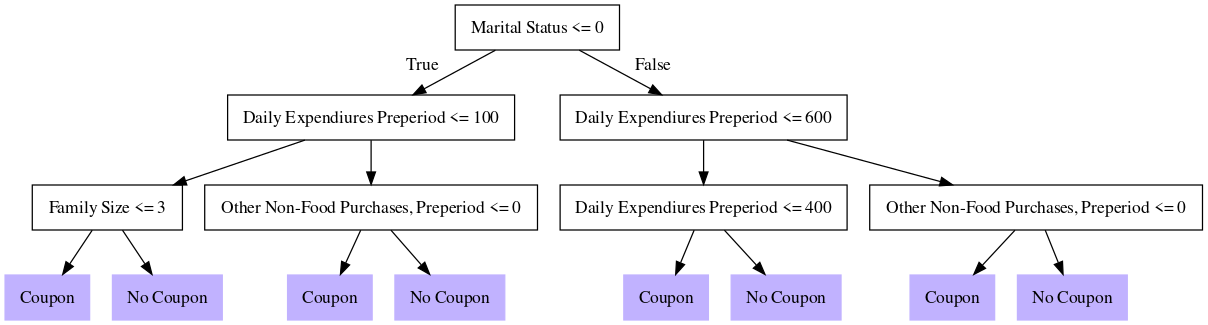} \\
				\textbf{(c)}  & \textbf{(d)} \\[25pt]
		\end{tabular}}
		\begin{tabular}{ccc}
			&\includegraphics[width=.6\textwidth]{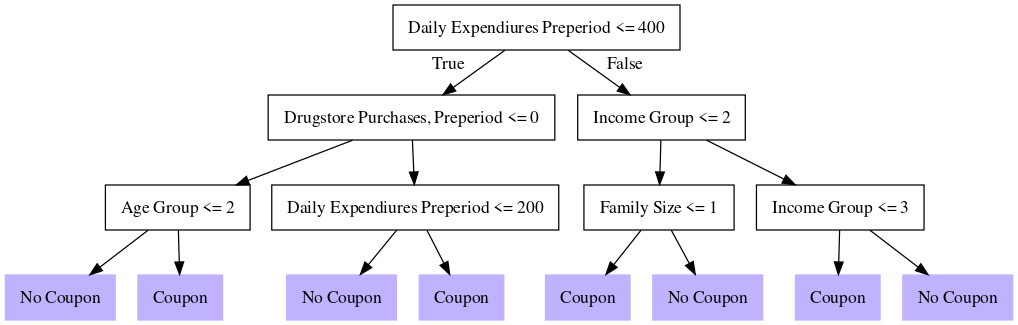}&\\
			&\textbf{(e)}&\\
		\end{tabular}
		\caption[Depth-3 trees for coupon provision, estimated in reduced dataset]{Depth-3 trees for coupons applicable to (a) ready-to-eat food, (b) meat and seafood, (c) other food, (d) drugstore products and (e) other non-food products, estimated in reduced dataset.}
		\label{figure:trees_robustness}
	\end{figure}
\end{appendix}
\end{document}